\documentclass[letter, a4paper]{IEEEtran}
\usepackage{amsmath,amsfonts}
\usepackage{algorithm}
\usepackage{array}
\usepackage{textcomp}
\usepackage{stfloats}
\usepackage{url}
\usepackage{verbatim}
\usepackage{graphicx}
\usepackage{cite}
\usepackage{float} 
\usepackage{authblk}
\usepackage{lipsum}
\usepackage[mathscr]{euscript}
\usepackage{framed}
\usepackage{algpseudocode}
\usepackage{bm}
\usepackage{stfloats}  
\usepackage{setspace}
\usepackage{balance}
\usepackage{amsthm} 

\usepackage{etoolbox}

\usepackage[caption=false,font=footnotesize]{subfig} 

\usepackage{graphicx,amsmath,amssymb,amsfonts}
\allowdisplaybreaks[3]
\usepackage{tikz}
\usetikzlibrary{arrows.meta,backgrounds,calc,decorations.pathreplacing,fit,positioning,shapes.geometric}
\usepackage{hyperref}
\hypersetup{colorlinks=true}   
\usepackage{xcolor}
\usepackage[T1]{fontenc}   
\newtheorem{remark}{\textbf{Remark}}
\newtheorem{lemma}{\textbf{Lemma}}

\newtheorem{corollary}{\textbf{Corollary}}

\newtheorem{proposition}{\textbf{Proposition}}

\floatname{algorithm}{Algorithm}

\makeatletter
\renewcommand{\maketag@@@}[1]{\hbox{\m@th\normalsize\normalfont#1}}%
\makeatother

\makeatletter

\newcommand{\Rmnum}[1]{\expandafter\@slowromancap\romannumeral #1@}

\makeatother

\begin{document}
	\captionsetup{font={small}}
	\bstctlcite{reference:BSTcontrol}
	\title{\fontsize{19 pt}{\baselineskip}\selectfont Sparse Rotatable Arrays (SRAs) for Cell-Free MIMO: Network-Wide Design of Antenna Activation and Panel Orientation}
		\author{
		\fontsize{10 pt}{\baselineskip}\selectfont Ailing~Zheng, Qingqing~Wu, Xiyuan~Liu,   Wei~Feng, Wen~Chen, Guoying~Zhang
		\vspace{-10 mm}
		\thanks{A. Zheng, Q. Wu, X. Liu, W. Chen, and G. Zhang are with the Department of Electronic Engineering, Shanghai Jiao Tong University, Shanghai 200240, China (e-mail: (ailing.zheng, qingqingwu, xiyuanliu\_SJTU, wenchen, gy\_zhang)@sjtu.edu.cn).}
			\thanks{W. Feng is with the Department of Electronic
				Engineering, State Key Laboratory of Space Network
				and Communications, Tsinghua University,
				Beijing 100084, China
				(e-mail: fengwei@tsinghua.edu.cn).}
				
	}

		\vspace{-10 mm}
	
	\maketitle
		\vspace{-5 mm}
	
	\begin{abstract}
			This paper investigates a sparse rotatable array
			(SRA)-enabled cell-free multiuser
			multiple-input multiple-output (MIMO) system.
			Under a network-wide antenna activation budget,
			distributed access points (APs) select active
			antenna elements from their installed arrays
			and adjust panel orientations, enabling joint
			control of sparse array apertures and antenna
			directivity.
			To avoid frequent hardware reconfiguration
			in response to instantaneous channel variations,
			we formulate a two-timescale stochastic
			optimization problem to maximize the expected
			minimum-user rate.
			Long-term antenna activation and panel
			orientations are optimized based on user
			spatial distributions and statistical channel
			state information (CSI), while short-term
			precoders adapt to instantaneous CSI.
			Under line-of-sight (LoS) propagation, we derive the optimal single-user antenna allocation, characterize the two-user max-min signal-to-interference-plus-noise ratio (SINR) and establish conditions for discrete activation and panel rotation to attain the continuous-position benchmark, and obtain a closed-form multi-AP zero-forcing (ZF) allocation benchmark under a continuous antenna-count relaxation.
		For the general mixed-integer nonconvex stochastic problem, we develop an alternating
			algorithm based on sample-average approximation
			(SAA), integrating uplink-downlink (UL-DL)
			duality for short-term precoding,
			weighted minimum mean-square error (WMMSE)
			surrogates for panel-orientation optimization,
			and multi-start greedy search for antenna
			activation.
			Numerical results demonstrate that the proposed design outperforms the fixed-array benchmark, improves all-user service reliability with fewer active antennas, and adapts effectively to nonuniform user spatial distributions.
	\end{abstract}

		\vspace{-2mm}

\vspace{-4mm}
\section{Introduction}
\vspace{-2mm}
\label{sec:introduction}
Future wireless networks are expected to support
massive device connectivity and increasingly diverse
services with stringent requirements on data rates,
reliability, and service quality \cite{Wang6G2023}.
Meeting these demands calls for large-scale antenna
deployments to provide sufficient spatial degrees of
freedom (DoFs) and accommodate geographically distributed
users. However, conventional cell-centric networks
remain susceptible to inter-cell interference and
substantial variations in link quality, particularly
for cell-edge users. To overcome these limitations,
cell-free massive multiple-input multiple-output
(MIMO) has emerged as a promising network
architecture
\cite{ngo2017cellfree,interdonato2019ubiquitous,demir2021foundations}.
By allowing geographically distributed access points
(APs) to cooperatively serve users over the same
time-frequency resources, cell-free transmission
exploits macro-diversity and joint signal
processing to improve coverage and provide more
uniform service quality.
Building on these advantages, extensive studies have investigated cooperative
precoding \cite{miretti2022team}
and resource allocation for spectral efficiency
\cite{Feng2022SE}, user fairness
\cite{farooq2021utility}, and energy efficiency
\cite{Papazafeiropoulos2021EE}.

Despite these advances, conventional cell-free systems primarily improve transmission through baseband signal processing, while their physical antenna configurations remain largely fixed. Under nonuniform user distributions, fixed AP-wise antenna allocation may lead to inefficient utilization of limited radio-frequency (RF) chains, while fixed array geometries and orientations constrain the achievable channel strength and spatial separability. Recent studies have investigated adaptive RF-chain activation and antenna selection to improve hardware efficiency and spectral efficiency~\cite{nguyen2022rfactivation,wang2026antennaselection}. However, antenna selection on a fixed panel can only modify the active array layout, while inter-user channel correlations remain constrained by panel orientation and element directional gains remain unchanged. This motivates transmitter-side array reconfiguration through network-wide antenna activation and panel rotation to jointly adapt channel strength and spatial separability.

Sparse rotatable arrays (SRAs) offer a promising architecture for such spatial reconfiguration~\cite{zheng2026sra}.  The essential idea is to activate a subset of antenna elements at predefined positions while exploiting orientation-dependent antenna gains. Building on this principle, we consider an SRA-enabled cell-free network where each AP is equipped with a rotatable panel hosting multiple pre-installed antenna elements. Specifically, antenna activation determines the number and spatial arrangement of active elements, while panel rotation adjusts their global positions and boresight direction, thereby modifying propagation phases and directional gains. Furthermore, under a network-wide activation budget, active antennas can be flexibly allocated across distributed APs according to their service requirements. Together, these mechanisms allow the network to adapt channel strength and inter-user correlation to different user distributions without increasing the total number of active antennas.

The above capability originates from a distinct
spatial reconfiguration principle that differentiates
the proposed SRA from existing flexible antenna
architectures.
Movable antennas (MAs) and fluid antennas
(FAs) enable position reconfiguration within
prescribed regions, thereby modifying
propagation phases and channel correlations
\cite{Zhu2024MA,Wong2020,Zhu2024CM}. Rotatable antennas (RAs)
instead exploit orientation adjustment to
control directional gains while preserving
the original array geometry \cite{ZhengRA2026,Zheng2026}.
More flexible configurations have been
investigated through six-dimensional movable
antennas (6DMAs), which allow both antenna
positions and array orientations to be
adjusted~\cite{Shao2025-1,Shao20256DMA}. In contrast, the SRA
architecture employs
a different configuration strategy based on
discrete antenna activation and panel-wise
rotation.
This design avoids individual antenna movement and element-wise orientation control while retaining flexibility in array geometry and antenna directivity.

Recent studies have extended antenna reconfiguration to distributed multi-AP networks. In RA-assisted cell-free downlink systems, antenna orientations have been jointly optimized with AP-user association or transmit precoding to improve sum-rate and fairness performance~\cite{pan2025racellfree,peng2026racellfree}. In~\cite{shi2025cellfree6dma}, the orientations of 6DMA surfaces were optimized according to user distributions to enhance the average sum rate, while~\cite{pi2026multiap6dma} investigated joint position and orientation optimization for multi-AP uplink coordination. More recently, a two-timescale framework was proposed for 6DMA-aided cell-free uplink systems, jointly optimizing long-term antenna positions and orientations with short-term receive beamforming~\cite{zhang2026twotimescale6dma}.
Despite these advances, the joint design of network-wide antenna activation and panel rotation in SRA-enabled cell-free networks remains insufficiently explored, particularly for downlink max-min fairness under limited antenna resources.
 In such networks, active user sets and instantaneous channels vary more rapidly than their underlying spatial and channel statistics. Frequent adjustment of antenna activation and panel orientations would incur considerable hardware control and coordination overhead, whereas digital precoding can readily adapt to instantaneous CSI. This motivates a two-timescale stochastic design that optimizes long-term antenna configurations based on user spatial and channel statistics, while adapting short-term downlink precoders to instantaneous channel realizations.

While the two-timescale framework avoids frequent
physical reconfiguration, the resulting joint design
remains challenging.  The discrete antenna activation determines the active array layouts, whereas continuous panel orientations affect both propagation phases and directional gains. Their joint influence on the effective channels leads to a mixed-integer nonconvex problem, in which changes that strengthen desired signals may also increase multiuser interference. Moreover, the network-wide activation budget couples the antenna configurations across APs, requiring local array designs to account for their roles in cooperative transmission. Furthermore, the long-term objective depends on random user sets and channel realizations, requiring repeated optimization of short-term precoders across multiple scenarios and making the configuration search computationally demanding.

To address the above challenges, this paper investigates an SRA-enabled cell-free downlink system. The main contributions are summarized as follows.

\begin{itemize}
	
	\item We investigate an SRA architecture for cell-free networks, enabling joint optimization of antenna activation across distributed APs and panel orientations. Under a network-wide activation budget and per-AP power constraints, we formulate a two-timescale stochastic optimization problem to maximize the expected minimum-user rate through  long-term antenna configurations and short-term downlink precoding.

	\item We characterize the performance mechanisms of antenna activation and panel rotation under line-of-sight (LoS) propagation. We first derive the optimal single-user signal-to-noise ratio (SNR) and antenna allocation across APs. For two-user transmission, we characterize the optimal max–min signal-to-interference-plus-noise ratio (SINR) under linear precoding and establish conditions under which discrete antenna activation and panel rotation attain the continuous-position benchmark by balancing channel strength and orthogonality. For multi-AP multiuser transmission, we characterize cross-AP correlation cancellation and derive the optimal zero-forcing (ZF) max–min SINR and closed-form antenna allocation under a fixed-structure continuous-count relaxation, providing guidelines for antenna activation initialization.

	\item We develop a sample average approximation (SAA)-based alternating optimization algorithm for the resulting mixed-integer nonconvex problem. Short-term max--min precoding is optimized via uplink-downlink (UL-DL) duality, while long-term panel orientations and antenna activation are updated through weighted minimum mean-square error (WMMSE)-based surrogate optimization and multi-start greedy search, respectively.
	
\item Numerical results validate the effectiveness of the proposed design. The proposed scheme achieves a 60.8\% higher average minimum-user rate than the fixed arrays and substantially improves all-user service reliability with fewer active antennas. Furthermore, joint antenna activation and panel rotation outperform their individual optimization, while adaptive antenna configurations improve performance across diverse user spatial distributions.
	
\end{itemize}

The remainder of this paper is organized as follows.
Section~\ref{System Model} introduces the system model
and problem formulation.
Section~\ref{Performance Analysis} investigates
the effects of antenna activation and panel rotation
on channel properties and multi-AP antenna allocation.
Section~\ref{Algorithm Design} develops the proposed
two-timescale optimization algorithm.
Section~\ref{subsec:simulation_setup} presents numerical results, and
Section~\ref{Conclusion} concludes the paper.

\vspace{-3mm}
		\section{System Model}  \label{System Model} 
		\vspace{-1mm}

\begin{figure}[t]
	\centering
	\includegraphics[width=0.45 \textwidth]{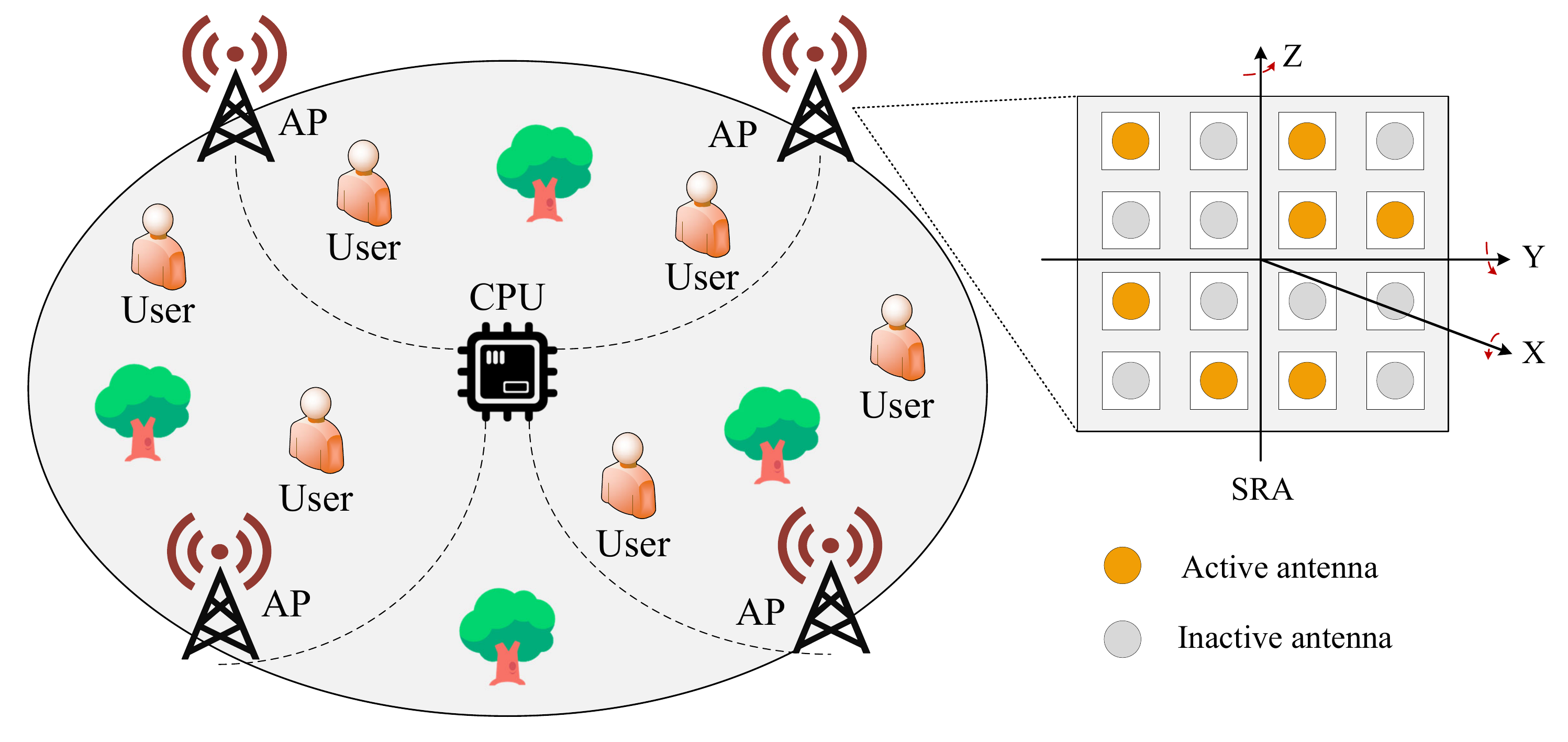}
	\caption{The considered downlink cell-free network with network-wide antenna activation and three-dimensional rotatable panels.}
	\label{Fig1}
	\vspace{-7mm}
\end{figure}
We consider a downlink cell-free network with $B$
distributed APs jointly serving a random number
$K$ of single-antenna users, as illustrated
in Fig.~\ref{Fig1}.
The APs and users are indexed by
$\mathcal B\triangleq\{1,\ldots,B\}$ and
$\mathcal K\triangleq\{1,\ldots,K\}$, respectively. Each user is equipped with a fixed isotropic antenna.
Each AP $b\in\mathcal B$ is connected to a central processing unit (CPU) via a dedicated fronthaul link and equipped with a rotatable panel comprising $\bar N_b$ pre-installed directional antennas, of which a subset can be activated in each long-term configuration interval.

We establish a global Cartesian coordinate system (GCS)
$x_g-y_g-z_g$.
The installation position of AP $b$ is fixed at
$\mathbf p_b=[x_b,y_b,z_b]^{\mathrm T}\in\mathbb R^3$,
while user $k\in\mathcal K$ is located at
$\mathbf q_k=[x_k,y_k,z_k]^{\mathrm T}\in\mathbb R^3$.
For each AP $b$, we define a local Cartesian coordinate
system (LCS) centered at $\mathbf p_b$,
with orthonormal basis vectors
$\mathbf e_x$, $\mathbf e_y$, and $\mathbf e_z$.
The antenna panel lies in the local
$\mathbf e_y$--$\mathbf e_z$ plane and contains
a uniform planar array
of $\bar N_b=N_{b,1}N_{b,2}$ antennas,
with inter-element spacing $d=\lambda/2$, where \(\lambda\) denotes the carrier wavelength.
For $i=1,\ldots,N_{b,1}$ and
$j=1,\ldots,N_{b,2}$,
the antenna index is $n=i+(j-1)N_{b,1}$,
and its local position is
	\setlength\abovedisplayskip{1pt}
\setlength\belowdisplayskip{1pt}
\begin{equation}
	\mathbf t_{b,n}
	=
	\left(i-\frac{N_{b,1}+1}{2}\right)d\,\mathbf e_y
	+
	\left(j-\frac{N_{b,2}+1}{2}\right)d\,\mathbf e_z.
	\label{eq:installed_antenna_position}
\end{equation}
	Let $a_{b,n}\in\{0,1\}$ indicate whether
	antenna $n$ at AP $b$ is activated.
	The activation vector of AP $b$ is defined as
	$\mathbf a_b\triangleq
	[a_{b,1},\ldots,a_{b,\bar N_b}]^{\mathrm T}$.
	The number of active antennas at AP $b$ is
	$n_b\triangleq\sum_{n=1}^{\bar N_b}a_{b,n}$,
	subject to $n_b\leq N_b^{\mathrm{RF}}$ and $
		\sum_{b\in\mathcal B}n_b=N_{\mathrm{act}}$,
	where $N_b^{\mathrm{RF}}\leq\bar N_b$
	denotes the number of available RF chains at AP $b$,
	and $N_{\mathrm{act}}$ is the fixed network-wide
	activation budget.
	Let $\mathbf R_b^{(0)}\in\mathrm{SO}(3)$ denote the reference
	orientation of panel $b$, where $\mathrm{SO}(3)$ denotes the three-dimensional special
	orthogonal group. Parameterized by the rotation angles
	$(\alpha_b,\beta_b,\gamma_b)$, the rotation matrix mapping
	local coordinates to the GCS is
	\begin{equation}
		\begin{aligned}
		\!\!\!\!	\mathbf R_b
			& \!=\! \mathbf R_b^{(0)}
			\mathbf R_x(\alpha_b)
			\mathbf R_y(\beta_b)
			\mathbf R_z(\gamma_b) \\
			&\!=\! \mathbf R_b^{(0)}
			\begin{bmatrix}
				c_\beta c_\gamma
				& -c_\beta s_\gamma
				& s_\beta \\
				c_\alpha s_\gamma+s_\alpha s_\beta c_\gamma
				& c_\alpha c_\gamma-s_\alpha s_\beta s_\gamma
				& -s_\alpha c_\beta \\
				s_\alpha s_\gamma-c_\alpha s_\beta c_\gamma
				& s_\alpha c_\gamma+c_\alpha s_\beta s_\gamma
				& c_\alpha c_\beta
			\end{bmatrix},
		\end{aligned}
		\label{eq:panel_rotation_parameterization}
	\end{equation}
	where $c_\xi\triangleq\cos\xi$ and $s_\xi\triangleq\sin\xi$,
	with the panel index $b$ omitted from the angles for brevity.
		The feasible rotation-matrix set of panel $b$ is
	\begin{equation}
		\begin{aligned}
			\mathcal R_b\triangleq
			\bigl\{
			&\mathbf R_b
			\big|\
			\alpha\in[\alpha_b^{\min},\alpha_b^{\max}],
			\beta\in[\beta_b^{\min},\beta_b^{\max}],\\
			&\gamma\in[\gamma_b^{\min},\gamma_b^{\max}]
			\bigr\}.
		\end{aligned}
		\label{eq:panel_angle_limits}
	\end{equation}
	Since the panel rotates about its fixed center,
	the global position of installed antenna $n$ is
$\mathbf r_{b,n}
		=
		\mathbf p_b+\mathbf R_b\mathbf t_{b,n}$.

	\vspace{-5mm}
\subsection{Channel Model and Long-Term User Distribution}
\vspace{-2mm}
\label{subsec:downlink_channel}

We consider a narrowband block-fading downlink
channel consisting of a LoS path
and multiple non-LoS (NLoS) paths.
We assume far-field propagation for each
AP-user link.
For a user at location $\mathbf q$ and AP $b$,
we define
\begin{equation}
	d_b(\mathbf q)
	\triangleq \|\mathbf q-\mathbf p_b\|_2,
	\boldsymbol\rho_{b,0}(\mathbf q)
	\triangleq
	{(\mathbf q-\mathbf p_b)}/{d_b(\mathbf q)},
	\label{eq:los_departure_direction}
\end{equation}
where $\boldsymbol\rho_{b,0}(\mathbf q)$
denotes the LoS departure direction.
Let $L_b(\mathbf q)$ denote the number of
NLoS paths, indexed by
$l=1,\ldots,L_b(\mathbf q)$, with the
LoS path indexed by $l=0$.
For each NLoS path,
$\boldsymbol\rho_{b,l}(\mathbf q)\in\mathbb R^3$
denotes its unit departure direction in the GCS.
For each path $l=0,\ldots,L_b(\mathbf q)$,
$\beta_{b,l}(\mathbf q)\geq 0$ denotes
the average path power before accounting
for directional element gain.
The total average path power is modeled as
\begin{equation}
	\beta_{\mathrm{tot},b}(\mathbf q)
	=\sum\nolimits_{l=0}^{L_b(\mathbf q)}
	\beta_{b,l}(\mathbf q)
	=\beta_{\mathrm{ref}}
	\left(\frac{d_b(\mathbf q)}{d_0}\right)^{
		-\nu_{\mathrm{PL}}},
	\label{eq:large_scale_path_loss}
\end{equation}
where $d_0$ is the reference distance,
$\beta_{\mathrm{ref}}
=(\lambda/(4\pi d_0))^2$
is the reference channel power gain,
and $\nu_{\mathrm{PL}}$ is the path-loss
exponent.
For a given location, these propagation
parameters remain approximately constant
over a long-term configuration interval.
For user $k$ at location $\mathbf q_k$,
let
$d_{k,b}=d_b(\mathbf q_k)$ and
$\boldsymbol\rho_{k,b,l}
=\boldsymbol\rho_{b,l}(\mathbf q_k)$
for $l=0,\ldots,L_b(\mathbf q_k)$.
The departure direction of path $l$ in
the LCS of AP $b$ is
$\widetilde{\boldsymbol\rho}_{k,b,l}
=\mathbf R_b^{\mathrm T}
\boldsymbol\rho_{k,b,l}$.
The response of all $\bar N_b$ installed
antennas to path $l$ is
\begin{equation}
	\bar{\mathbf a}_{k,b,l}(\mathbf R_b)
	\triangleq
	\left[
	\exp\!\left(
	j\frac{2\pi}{\lambda}
	\boldsymbol\rho_{k,b,l}^{\mathrm T}
	\mathbf R_b\mathbf t_{b,n}
	\right)
	\right]_{n=1}^{\bar N_b}
	\in\mathbb C^{\bar N_b}.
	\label{eq:full_array_response}
\end{equation}
All directional antennas on the panel share
the local boresight $\mathbf e_x$.
The angle between this boresight and the
departure direction of path $l$ is
$\varepsilon_{k,b,l}
=\arccos(
\mathbf e_x^{\mathrm T}
\widetilde{\boldsymbol\rho}_{k,b,l})$.
We adopt the directional element power
pattern~\cite{AilngRASecure2026,balanis2016antenna}
\begin{equation}
	G(\varepsilon)=
	\begin{cases}
		2(2p+1)\cos^{2p}\varepsilon,
		&0\leq\varepsilon\leq\pi/2,\\
		0,
		&\pi/2<\varepsilon\leq\pi,
	\end{cases}
	\label{eq:directional_element_pattern}
\end{equation}
where $p \ge 0$ controls the directivity of the antenna element.
Accordingly, the field-amplitude gain
of path $l$ is
$A_{k,b,l}(\mathbf R_b)
\triangleq\sqrt{G(\varepsilon_{k,b,l})}$.
The LoS path coefficient is
\begin{equation}
	\alpha_{k,b,0}
	=\sqrt{\beta_{b,0}(\mathbf q_k)}
	\exp\!\left(
	-j\frac{2\pi d_{k,b}}{\lambda}
	\right).
	\label{eq:los_path_coefficient}
\end{equation}
For each NLoS path $l\geq1$, we have
\begin{equation}
	\alpha_{k,b,l}
	=\sqrt{\beta_{b,l}(\mathbf q_k)}
	z_{k,b,l},
	z_{k,b,l}\sim\mathcal{CN}(0,1),
	\label{eq:nlos_path_coefficient}
\end{equation}
where $z_{k,b,l}$ is the normalized
small-scale fading coefficient.
	The physical channel vector of all installed antennas
at AP \(b\) can then be written as
\begin{equation}
	\bar{\mathbf g}_{k,b}
	=\sum\nolimits_{l=0}^{L_b(\mathbf q_k)}
	\alpha_{k,b,l}
	A_{k,b,l}(\mathbf R_b)
	\bar{\mathbf a}_{k,b,l}(\mathbf R_b)
	\in\mathbb C^{\bar N_b}.
	\label{eq:full_ap_user_channel}
\end{equation}  
Let
$\mathbf D_b\triangleq
\operatorname{diag}
(a_{b,1},\ldots,a_{b,\bar N_b})$.
The effective downlink channel vector is
$\mathbf g_{k,b}
\triangleq\mathbf D_b\bar{\mathbf g}_{k,b}
\in\mathbb C^{\bar N_b}$.

Let $\mathcal L\subseteq\mathbb R^3$ denote
a two-dimensional service region on the
horizontal user plane.
We model the active users as a nonhomogeneous
Poisson point process with spatial intensity
$\rho(\mathbf q)$.
The mean number of active users is
$\mu\triangleq
\int_{\mathcal L}\rho(\mathbf q)
\,\mathrm dA(\mathbf q)>0$,
where $\mathrm dA(\mathbf q)$ denotes the
area element.
Accordingly, the number $K$ of active
users satisfies
\begin{equation}
	\Pr(K=k)
	=\frac{\mu^k e^{-\mu}}{k!},
	 k=0,1,\ldots.
	\label{eq:poisson_user_number}
\end{equation}
Conditioned on $K=k$, their locations
are independently drawn from the density
\begin{equation}
	f(\mathbf q)
	=\frac{\rho(\mathbf q)}{\mu},
	 \mathbf q\in\mathcal L.
	\label{eq:user_location_density}
\end{equation}
We denote a user realization by
$\mathcal U=(K,\mathbf q_1,\ldots,\mathbf q_K)$.

\vspace{-4mm}
\subsection{Signal Model and Problem Formulation}
\vspace{-1mm}

We collect the panel rotation angles in
$\boldsymbol\vartheta\triangleq
(\alpha_1,\beta_1,\gamma_1,\ldots,
\alpha_B,\beta_B,\gamma_B)^{\mathrm T}$.
The feasible angle set is the closed convex box
\begin{equation}
\!\!\!\!\! \mathcal V\triangleq
\prod_{b=1}^{B}\left(
[\alpha_b^{\min},\alpha_b^{\max}]
\times[\beta_b^{\min},\beta_b^{\max}]
\times[\gamma_b^{\min},\gamma_b^{\max}]
\right).
\label{eq:feasible_angle_box}
\end{equation}
Each $\boldsymbol\vartheta\in\mathcal V$ induces panel rotations
$\mathbf R_b(\alpha_b,\beta_b,\gamma_b)\in\mathcal R_b$
as defined in \eqref{eq:panel_angle_limits}.
The long-term configuration consists of the antenna activation pattern $\mathbf a=\{\mathbf a_{b}\}_{b}$ and the panel orientations $\boldsymbol\vartheta$, denoted by
$\boldsymbol\Theta\triangleq(\mathbf a,\boldsymbol\vartheta)$.
For a realized channel \(\mathbf H=\{\mathbf g_{k,b}\}_{k,b}\), let \(\mathbf w_{b,j}\in\mathbb C^{\bar N_b}\) denote the precoder used by AP \(b\) for user \(j\), and let \(\mathbf W=\{\mathbf w_{b,j}\}_{b,j}\). With independent information symbols \(s_j\sim\mathcal{CN}(0,1)\), the received signal at user \(k\) is
\begin{equation}
	y_k=
	\sum\nolimits_{j=1}^{K}
	\left(
	\sum\nolimits_{b=1}^{B}
	\mathbf g_{k,b}^{\mathrm T}\mathbf w_{b,j}
	\right)s_j+z_k,
	 k=1,\ldots,K,
	\label{eq:dl_received_signal}
\end{equation}
where \(z_k\sim\mathcal{CN}(0,\sigma_k^2)\). Then, we have the instantaneous SINR of user \(k\)
\begin{equation}
	\gamma_k(\mathbf H,\mathbf W)
	=
	\frac{
		\left|
		\sum\nolimits_{b=1}^{B}
		\mathbf g_{k,b}^{\mathrm T}\mathbf w_{b,k}
		\right|^2
	}{
		\displaystyle
		\sum\nolimits_{{j=1,j\ne k}}^{K}
		\left|
		\sum\nolimits_{b=1}^{B}
		\mathbf g_{k,b}^{\mathrm T}\mathbf w_{b,j}
		\right|^2+\sigma_k^2
	}.
	\label{eq:instantaneous_sinr}
\end{equation}
The corresponding achievable rate is
$R_k(\mathbf H,\mathbf W)
	=
	\log_2\!\left(1+\gamma_k(\mathbf H,\mathbf W)\right)$.
For a nonempty user realization
$\mathcal U$, we define the feasible precoder set as
\begin{equation}
	\begin{aligned}
	\!\!\!\!\!\!	\mathcal W(\mathcal U;\mathbf a)
	\!	\triangleq\!
		\big\{ \mathbf W  \big| 
		\sum\nolimits_{j=1}^{K}\! \|\mathbf w_{b,j}\|_2^2
		\leq \! P_b,
		\mathbf w_{b,j}
		\!=\!\mathbf D_b\mathbf w_{b,j}
		\big\},
	\end{aligned}
	\label{eq:precoding_feasible_set}
\end{equation}
where $P_b$ is the total transmit-power budget of AP $b$.
For each realization $(\mathcal U,\mathbf H)$,
the optimal minimum instantaneous user rate is
\begin{equation}
	V(\mathcal U,\mathbf H;\boldsymbol\Theta)
	\triangleq
	\max_{\mathbf W\in
		\mathcal W(\mathcal U;\mathbf a)}
	\min_{1\leq k\leq K}
	R_k(\mathbf H,\mathbf W).
	\label{eq:instantaneous_max_min_value}
\end{equation}
The corresponding long-term performance is
\begin{equation}
	F(\boldsymbol\Theta)
	\triangleq
	\mathbb E_{\mathcal U,\mathbf H
		\mid K\geq 1;\boldsymbol\Theta}
	\left[
	V(\mathcal U,\mathbf H;\boldsymbol\Theta)
	\right],
	\label{eq:long_term_objective}
\end{equation}
where the expectation accounts for random
user realizations and their corresponding
channels under the long-term configuration
$\boldsymbol\Theta$.
Accordingly, the joint two-timescale
design problem is formulated as
\begin{subequations}
	\label{eq:long_term_design_problem}
	\begin{align}
		\max_{\mathbf a,\boldsymbol\vartheta}
		\quad&
			F(\boldsymbol\Theta)
		\label{eq:long_term_design_objective}
		\\
		\mathrm{s.t.}\quad&
		a_{b,n}\in\{0,1\},	\forall b,n,
		\sum\nolimits_{n=1}^{\bar N_b}a_{b,n}
		\leq N_b^{\mathrm{RF}},
		\forall b,
		\label{eq:long_term_rf_constraint}
		\\
		&
		\sum\nolimits_{b=1}^{B}
		\sum\nolimits_{n=1}^{\bar N_b}a_{b,n}
		=N_{\mathrm{act}},
		\boldsymbol\vartheta\in\mathcal V.
		\label{eq:long_term_orientation_constraint}
	\end{align}
\end{subequations}
The resulting problem is mixed-integer and
nonconvex due to the coupling between binary
antenna activation and continuous panel
orientations through the effective channels.
Moreover, evaluating $F(\boldsymbol\Theta)$
requires solving a short-term precoding problem
for each user-channel realization and
averaging the resulting optimal rates.
To address these challenges, we employ
SAA and
develop an alternating optimization algorithm
for antenna activation and panel orientations,
with scenario-specific precoders reoptimized
at each iteration.

\vspace{-3mm}
\section{Performance Analysis}
\label{Performance Analysis}
\vspace{-1mm}

This section investigates the complementary roles
of antenna activation and panel rotation in shaping
channel strength and spatial separability under
LoS propagation.
We first characterize the optimal single-user SNR
and antenna allocation across APs, then examine
the tradeoff between channel strength and
orthogonality in two-user transmission, and
finally analyze distributed channel separation
and ZF antenna allocation under a fixed-structure
continuous-count model.

\vspace{-3mm}
\subsection{Optimal Antenna Allocation for Single-User Transmission}
\label{sec:single_user_benchmark}
\vspace{-1mm}
We consider a single user at a fixed location $\mathbf q$ under
LoS propagation. Let
$\sigma^2>0$ denote its noise power. The per-antenna LoS channel power gain
at AP $b$ is defined as
\begin{equation}
	\eta_b(\mathbf R_b)
	\triangleq
	\beta_{b,0}(\mathbf q)
	G\bigl(\varepsilon_{b,0}(\mathbf q,\mathbf R_b)\bigr).
	\label{eq:su_effective_path_power}
\end{equation}
Under the far-field LoS model, the channel at AP $b$
satisfies $\|\mathbf g_b\|_2^2=n_b\eta_b(\mathbf R_b)$,
where $n_b=\sum_{n=1}^{\bar N_b}a_{b,n}$.

\begin{proposition}
	\label{prop:su_optimal_snr}
\upshape	For any feasible activation pattern and panel orientations,
	the optimal single-user SNR under coherent joint transmission
	and per-AP power constraints is
	\begin{equation}
		\gamma_{\mathrm{SU}}^\star(\mathbf n,\mathbf R)
		=
		\frac{1}{\sigma^2}
		\left(
		\sum\nolimits_{b=1}^{B}
		\sqrt{P_b\eta_b(\mathbf R_b)n_b}
		\right)^2.
		\label{eq:su_optimal_snr}
	\end{equation}
\end{proposition}

\begin{proof}
	For any feasible precoders,
	\begin{align}
	\!\!\!\!	\left| \sum\nolimits_{b=1}^{B}\!\! \mathbf g_b^{\mathrm T}\mathbf w_b \right|
	\!\!	\leq \!\!
		\sum\nolimits_{b=1}^{B} \! \|\mathbf g_b\|_2\|\mathbf w_b\|_2
	\!	\leq \!
		\sum\nolimits_{b=1}^{B} \!\! \sqrt{P_b}\|\mathbf g_b\|_2.
	\end{align}
	Both inequalities are attained by
	$\mathbf w_b^\star=\sqrt{P_b}\mathbf g_b^\ast/\|\mathbf g_b\|_2$
	for $\|\mathbf g_b\|_2>0$, and $\mathbf w_b^\star=\mathbf0$
	otherwise. These precoders satisfy
	$\mathbf w_b^\star=\mathbf D_b\mathbf w_b^\star$
	and align all nonzero received contributions in phase.
	Substituting $\|\mathbf g_b\|_2^2=n_b\eta_b(\mathbf R_b)$
	proves \eqref{eq:su_optimal_snr}.
\end{proof}

For fixed panel orientations, maximizing the SNR is
equivalent to the integer allocation problem
\begin{equation}
	\max_{\mathbf n\in\mathcal N_{\mathrm{int}}}
	\sum\nolimits_{b=1}^{B}c_b\sqrt{n_b},
	c_b\triangleq\sqrt{P_b\eta_b(\mathbf R_b)},
	\label{eq:su_integer_allocation}
\end{equation}
where
\begin{equation}
	\mathcal N_{\mathrm{int}}
	\triangleq
	\left\{
	\mathbf n\in\mathbb Z_{\geq0}^{B}\,\middle|\,
	n_b\leq N_b^{\mathrm{RF}},
	 \sum\nolimits_{b=1}^{B}n_b=N_{\mathrm{act}}
	\right\}.
	\label{eq:su_integer_feasible_set}
\end{equation}
The marginal amplitude gain of the $m$-th antenna at AP $b$ is
\begin{equation}
	\delta_b(m)
	\triangleq
	c_b\bigl(\sqrt m-\sqrt{m-1}\bigr),
	 1\leq m\leq N_b^{\mathrm{RF}}.
	\label{eq:su_marginal_gain}
\end{equation}
Since
$c_b\sqrt{n_b}=\sum_{m=1}^{n_b}\delta_b(m)$
and $\delta_b(m)$ is nonincreasing in $m$,
\eqref{eq:su_integer_allocation} admits an exact greedy solution.
Starting from $n_b=0$ for all APs, repeatedly assign one antenna to
\begin{equation}
	b^\star\in
	\arg\max_{b:\,n_b<N_b^{\mathrm{RF}}}
	\delta_b(n_b+1).
	\label{eq:su_greedy_allocation}
\end{equation}
The diminishing marginal gains $\delta_b(m)$ ensure
that \eqref{eq:su_greedy_allocation} globally solves
\eqref{eq:su_integer_allocation}.
Since the SNR increases monotonically with each
$\eta_b(\mathbf R_b)$ and the orientation constraints
are independent across APs, the joint optimum is
obtained by first choosing
$\mathbf R_b^\star\in
\arg\max_{\mathbf R_b\in\mathcal R_b}
G\bigl(\varepsilon_{b,0}(\mathbf q,\mathbf R_b)\bigr)$
for each AP, then applying
\eqref{eq:su_greedy_allocation}.
If boresight alignment is feasible,
$\mathbf R_b^\star\mathbf e_x=\boldsymbol\rho_{b,0}$
achieves the maximum directional gain.
\begin{remark}
	\upshape
	Under far-field LoS propagation, the optimal
	single-user SNR depends only on AP-wise antenna
	counts and boresight directions. Position selection
	and in-plane rotation instead affect multiuser
	separation, while activating antennas at additional
	APs makes their transmit-power budgets available.
\end{remark}

\vspace{-4mm}
\subsection{Channel Strength and Spatial Separability via Antenna Activation and Panel Rotation}
\label{subsec:III-B}
\vspace{-1mm}

While antenna positions do not affect the
single-user SNR under far-field LoS propagation,
they become crucial in multiuser transmission
by controlling inter-user channel correlation.
We consider a single AP serving two users
to characterize the tradeoff between
channel strength and spatial separability.

\subsubsection{Two-User LoS Geometry and SINR Characterization}
Let $P>0$ be the total transmit power, and let
$\boldsymbol{\rho}_k$ and $\beta_k>0$ denote the
departure direction and path gain of user $k$.
Define $\Delta\boldsymbol\rho=
\boldsymbol\rho_1-\boldsymbol\rho_2$.
Two antennas are activated on a rectangular
panel with separation $\mathbf d=\mathbf t_2-\mathbf t_1$.
The boresight is $\mathbf f=\mathbf R\mathbf e_x$,
and the per-antenna gain is
$\eta_k(\mathbf R)=
\beta_kG(\arccos(\boldsymbol\rho_k^{\mathrm T}\mathbf f))$.
The channel strength and squared normalized
correlation are $s_k=\|\mathbf g_k\|_2^2
=2\eta_k(\mathbf R)$ and

\begin{equation}
	r(\mathbf d,\mathbf R)
	=\frac{|\mathbf g_1^{\mathrm T}\mathbf g_2^\ast|^2}{s_1s_2}
	=\cos^2\left(
	\frac{\pi}{\lambda}
	\Delta\boldsymbol\rho^{\mathrm T}\mathbf R\mathbf d
	\right).
	\label{eq:b_two_user_correlation}
\end{equation}
	Equation~\eqref{eq:b_two_user_correlation} reveals
	that antenna position adjustment and panel rotation
	control channel correlation by changing the separation
	$\mathbf d$ and its projection through $\mathbf R$,
	respectively.
	For nonzero LoS channels, orthogonality is achieved
	when $\Delta\boldsymbol\rho^{\mathrm T}\mathbf R\mathbf d
	=(m+1/2)\lambda$, $m\in\mathbb Z$, independently
	of the directivity parameter $p$.
We define
$x=P\bigl(\sigma_1^2/s_1+\sigma_2^2/s_2\bigr)^{-1}$,
the optimal common SINR for orthogonal user channels
with the given strengths.

\begin{lemma}
	\label{lem:twouser_sinr}
	\upshape
	For two user channels under transmit power
	$P$, the optimal max-min SINR achievable by linear
	precoding is
	\begin{equation}
		\tau_2^\star=T(x,r)
		\triangleq x-\frac{rx^2}{1+x}.
		\label{eq:b_two_user_sinr}
	\end{equation}
	For $x>0$, $T(x,r)$ is strictly increasing in $x$
	and strictly decreasing in $r$, with $T(x,0)=x$.
\end{lemma}
\begin{IEEEproof}
	See Appendix~\ref{app:b_two_user_sinr}.
\end{IEEEproof}
Lemma \ref{lem:twouser_sinr} reveals that maximizing the minimum-user SINR requires balancing channel strength and correlation. Under far-field LoS propagation, antenna position selection controls channel correlation, whereas panel rotation affects both channel strength and correlation.

\subsubsection{Antenna Activation and Panel-Orientation Design}
We consider a rectangular panel of side lengths $L_1$ and $L_2$.  To identify the geometric limits,
we allow unrestricted panel rotation
$\mathbf R\in\mathrm{SO}(3)$.
For the continuous-position benchmark, the antennas can occupy
arbitrary panel positions. 
Let $\mathbf R_0$ denote a reference orientation
providing positive gains to both users.
For the fixed reference antenna pair at
positions $\mathbf t_1^{(0)}$ and
$\mathbf t_2^{(0)}$, we define
$\mathbf d_0=\mathbf t_2^{(0)}-\mathbf t_1^{(0)}$
and $\ell_0=\|\mathbf d_0\|_2$.

\textbf{1) Position-only design:}
With $\mathbf R=\mathbf R_0$, we define
$\mathbf g=\mathbf R_0^{\mathrm T}\Delta\boldsymbol\rho$
and $A_{\mathrm{mov}}=L_1|g_2|+L_2|g_3|$.
The projected separation spans
$[-A_{\mathrm{mov}},A_{\mathrm{mov}}]$.
The first correlation zero occurs at a
projected separation of $\lambda/2$.
Defining $x_0\triangleq x(\mathbf R_0)$,
we obtain from Lemma~\ref{lem:twouser_sinr}
\begin{equation}
	\tau_{\mathrm{mov}}^\star
	=T\left(
	x_0,
	\cos^2\left(
	\frac{\pi}{\lambda}
	\min\left\{A_{\mathrm{mov}},\frac{\lambda}{2}\right\}
	\right)
	\right),
	\label{eq:b_position_only}
\end{equation}
Thus, $\tau_{\mathrm{mov}}^\star=x_0$ if and only if
$A_{\mathrm{mov}}\geq\lambda/2$.

\textbf{2) Rotation-only and joint designs:}
For $\Delta\theta=\arccos(
\boldsymbol\rho_1^{\mathrm T}\boldsymbol\rho_2)\in(0,\pi)$,
we define
\begin{equation}
	a=\frac{\Delta\theta}{2},
	\mathbf m=\frac{\boldsymbol\rho_1+\boldsymbol\rho_2}
	{\|\boldsymbol\rho_1+\boldsymbol\rho_2\|_2},
	\mathbf v=\frac{\Delta\boldsymbol\rho}
	{\|\Delta\boldsymbol\rho\|_2}.
	\label{eq:b_rotation_geometry}
\end{equation}
Vectors $\mathbf m$ and $\mathbf v$ are orthonormal,
and user directions satisfy
\begin{equation}
\boldsymbol\rho_1=\cos a\,\mathbf m+\sin a\,\mathbf v,
\boldsymbol\rho_2=\cos a\,\mathbf m-\sin a\,\mathbf v,
\end{equation}
with $\Delta\boldsymbol\rho=2\sin a\,\mathbf v$.
An optimal boresight can be chosen in this plane,
since a boresight in the plane maximizes both directional
gains for a fixed component along $\mathbf v$ without
changing the attainable range of projected separations.
We therefore parameterize the boresight
by its angular offset $t$ from the
bisector $\mathbf m$ of the two user
directions as
\begin{equation}
	\mathbf f(t)=\cos t\,\mathbf m+\sin t\,\mathbf v,
	t\in\mathcal I
	\triangleq[-\pi/2+a,\pi/2-a].
	\label{eq:b_boresight_parameterization}
\end{equation}
The corresponding direction cosines are
$\boldsymbol\rho_1^{\mathrm T}\mathbf f(t)=\cos(a-t)$
and
$\boldsymbol\rho_2^{\mathrm T}\mathbf f(t)=\cos(a+t)$.
The interval $\mathcal I$ keeps both users in the
front hemisphere of the panel.
With $G_{\max}=2(2p+1)$, the directional pattern in
Section~\ref{subsec:downlink_channel} gives
\begin{align}
	s_1(t)\!=\!2\beta_1G_{\max}\cos^{2p}(a-t),
	s_2(t)\!=\!2\beta_2G_{\max}\cos^{2p}(a+t). \nonumber
\end{align}
Substituting these expressions into
$x=P(\sigma_1^2/s_1+\sigma_2^2/s_2)^{-1}$
and defining $c_k=\sigma_k^2/\beta_k$ yield
\begin{equation}
	x_2(t)=
	\frac{2PG_{\max}}
	{c_1\cos^{-2p}(a-t)+c_2\cos^{-2p}(a+t)}.
	\label{eq:b_rotation_strength}
\end{equation}
For $p>0$, at least one user has zero directional gain
at each endpoint of $\mathcal I$,
where we extend $x_2(t)$ by zero.

For a fixed boresight $\mathbf f(t)$,
let $\mathbf u=\mathbf R\mathbf d$ denote the global
antenna-separation vector, with
$\|\mathbf u\|_2=\ell$.
Since both antennas lie on the panel,
$\mathbf u^{\mathrm T}\mathbf f(t)=0$.
The maximum absolute projection of $\mathbf u$
onto $\Delta\boldsymbol\rho$ is therefore
\begin{align}
	A(t;\ell)
	&=\ell\left\|
	\bigl(\mathbf I-\mathbf f(t)\mathbf f(t)^{\mathrm T}\bigr)
	\Delta\boldsymbol\rho
	\right\|_2 \nonumber \\
	&=\ell\sqrt{
		\|\Delta\boldsymbol\rho\|_2^2
		-\bigl(\Delta\boldsymbol\rho^{\mathrm T}
		\mathbf f(t)\bigr)^2}
	=2\ell\sin a\cos t.
\end{align}
Under unrestricted panel rotation, rotating the panel
about $\mathbf f(t)$ makes
$\Delta\boldsymbol\rho^{\mathrm T}\mathbf u$
span $[-A(t;\ell),A(t;\ell)]$.
{Rotation about the fixed boresight $\mathbf f(t)$
	preserves both users' directional gains and channel strengths,
	while varying the projected antenna separation and hence
	their channel correlation.}
The correlation is minimized at the largest attainable
absolute projection below $\lambda/2$, or at the first
zero $\lambda/2$ when it is attainable.
Hence,
\begin{equation}
	r_{\min}(t;\ell)=
	\cos^2\left(
	\frac{\pi}{\lambda}
	\min\left\{2\ell\sin a\cos t,\frac{\lambda}{2}\right\}
	\right).
	\label{eq:b_rotation_correlation}
\end{equation}
Define
$\mathcal T(\ell)=
\max_{t\in\mathcal I}T\bigl(x_2(t),r_{\min}(t;\ell)\bigr)$.
Increasing $\ell$ enlarges the attainable projection
range and cannot increase $r_{\min}(t;\ell)$.
Since $T(x,r)$ decreases with $r$,
$\mathcal T(\ell)$ is nondecreasing in $\ell$.
The rotation-only design fixes the separation at $\ell_0$,
whereas continuous position adjustment permits a maximum
separation of
$\ell_{\max}=\sqrt{L_1^2+L_2^2}$,
attained by opposite panel corners.
Consequently, we obtain
\begin{equation}
	\tau_{\mathrm{rot}}^\star=\mathcal T(\ell_0),
	\tau_{\mathrm{mov+rot}}^\star=\mathcal T(\ell_{\max}).
	\label{eq:b_rotation_joint_sinr}
\end{equation}

By \eqref{eq:b_rotation_correlation},
orthogonality at $\mathbf f(t)$ is attainable
if and only if
$2\ell\sin a\cos t\geq\lambda/2$.
Since $\cos t\leq1$, with equality at
$t=0\in\mathcal I$,
the minimum separation permitting orthogonality is
\begin{equation}
	\ell_{\mathrm{crit}}
	\triangleq\frac{\lambda}{4\sin a}
	=\frac{\lambda}{2\|\Delta\boldsymbol\rho\|_2}.
	\label{eq:b_critical_separation}
\end{equation}
{
	The threshold $\ell_{\mathrm{crit}}$ characterizes
	the geometric feasibility of channel orthogonality,
	independently of the directivity parameter $p$.
}
Let $t_x=\arg\max_{t\in\mathcal I}x_2(t)$ and
$x_{\max}=x_2(t_x)$.
To attain maximum channel strength and orthogonality
simultaneously, the orthogonality condition must hold
at $t=t_x$, i.e.,
$\ell\cos t_x\geq\ell_{\mathrm{crit}}$.
The corresponding minimum separation is therefore
$\ell_{\mathrm{full}}
\triangleq\ell_{\mathrm{crit}}/\cos t_x$.
	Unlike $\ell_{\mathrm{crit}}$, $\ell_{\mathrm{full}}$
	generally depends on the directional pattern
	through $t_x$, reflecting the tradeoff between
	channel strength and orthogonality.

\begin{proposition}
	\label{prop:III-B-aperture-limits}
	\upshape
	Under the two-user LoS benchmark with unrestricted panel rotation and \(p>0\), the optimal max–min SINR \({\mathcal T(\ell)}\) is strictly increasing for \(0<\ell<\ell_{\mathrm{full}}\) and attains \(x_{\max}\) for \(\ell\ge\ell_{\mathrm{full}}\). Consequently, joint position and orientation adjustment strictly outperforms rotation alone if and only if \(\ell_{\max}>\ell_0\) and \(\ell_0<\ell_{\mathrm{full}}\).
\end{proposition}

\begin{proof}
	For $p>0$, the denominator in
	\eqref{eq:b_rotation_strength} is strictly convex,
	so $x_2(t)$ has a unique maximizer $t_x$ and
	strictly increases toward $t_x$.
	By Lemma~\ref{lem:twouser_sinr}, we have
	\begin{equation}
		T\bigl(x_2(t),r_{\min}(t;\ell)\bigr)
		\leq x_2(t)\leq x_{\max}.
	\end{equation}
	Equality throughout holds if and only if
	$t=t_x$ and $r_{\min}(t_x;\ell)=0$,
	equivalently, $\ell\geq\ell_{\mathrm{full}}$.
	Since $r_{\min}(t;\ell)$ is nonincreasing in $\ell$,
	$\mathcal T(\ell)$ is nondecreasing.
	For $0<\ell_1<\ell_2<\ell_{\mathrm{full}}$,
	let $t_1$ be an optimizer at $\ell_1$.
	If $r_{\min}(t_1;\ell_1)>0$, increasing $\ell$
	strictly reduces correlation and improves SINR.
	Otherwise, $t_1\neq t_x$, and
	$\ell_2\cos t_1>\ell_{\mathrm{crit}}$.
	A small shift toward $t_x$ thus preserves
	orthogonality while increasing $x_2(t)$.
	Hence, $\mathcal T(\ell_2)>\mathcal T(\ell_1)$.
	Finally, since
	$\tau_{\mathrm{rot}}^\star=\mathcal T(\ell_0)$
	and
	$\tau_{\mathrm{mov+rot}}^\star=\mathcal T(\ell_{\max})$,
	strict improvement holds if and only if
	$\ell_{\max}>\ell_0$ and
	$\ell_0<\ell_{\mathrm{full}}$.
\end{proof}

\begin{figure}[t]
	\centering
	\includegraphics[width=0.47\textwidth]{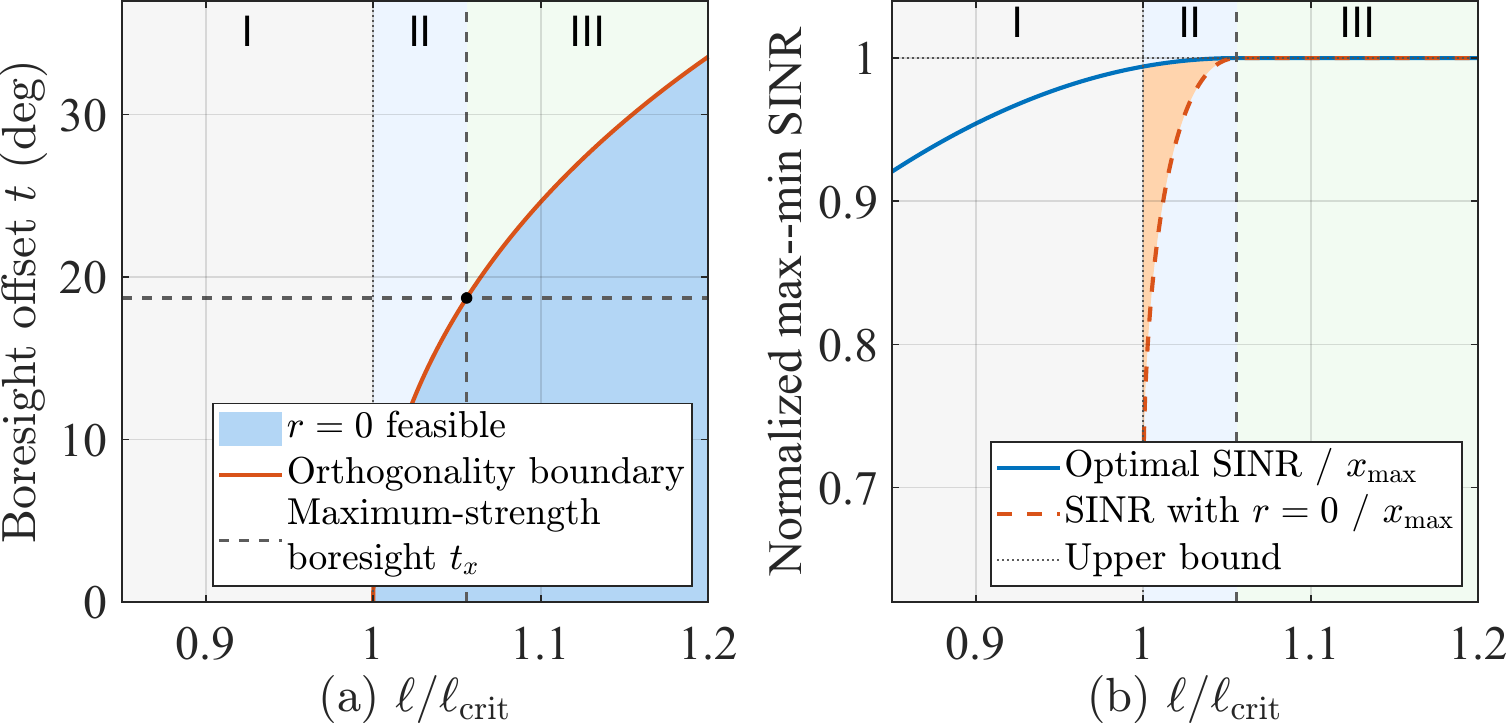}
	\caption{Aperture regimes in
		Proposition~\ref{prop:III-B-aperture-limits}.
		(a) Orthogonality-feasible boresight offsets.
		(b) Normalized optimal
		max-min SINRs.}
	\label{fig:aperture_thresholds}
	\vspace{-7mm}
\end{figure}
Proposition~\ref{prop:III-B-aperture-limits}
establishes strict SINR improvement with
antenna separation until saturation at
$\ell_{\mathrm{full}}$.
Fig.~\ref{fig:aperture_thresholds}(a)
shows that increasing $\ell$ expands the
boresight offsets permitting orthogonality.
For $\ell_{\mathrm{crit}}\leq\ell<
\ell_{\mathrm{full}}$, the optimum may
retain nonzero correlation to preserve
directional gain, as shown in
Fig.~\ref{fig:aperture_thresholds}(b).
At $\ell_{\mathrm{full}}$, maximum channel
strength and orthogonality become jointly
attainable. When $c_1=c_2$, we have $t_x=0$.
For $p=0$, $x_2(t)$ is independent of
$t$ within $\mathcal I$, and we choose
$t_x=0$ without loss of optimality.
In both cases, $\ell_{\mathrm{full}}
=\ell_{\mathrm{crit}}$, eliminating
the intermediate regime.

\textbf{3) Discrete antenna activation:}
For installed antenna positions
$\mathcal T_0 \!=\! \{\mathbf t_n^0\}_{n=1}^{\bar N}$,
the minimum correlation under
$\mathbf R_0$ is
\begin{equation}
	r_{\mathrm{sel}}^\star
	=\min_{1\leq m<n\leq\bar N}
	\cos^2\left(
	\frac{\pi}{\lambda}
	\Delta\boldsymbol\rho^{\mathrm T}\mathbf R_0
	(\mathbf t_m^0-\mathbf t_n^0)
	\right),
	\label{eq:III-B-discrete-correlation}
\end{equation}
yielding $\tau_{\mathrm{sel}}^\star
=T(x_0,r_{\mathrm{sel}}^\star)
\leq\tau_{\mathrm{mov}}^\star$.
The inequality can be strict because discrete
projected separations may miss the correlation zeros.
Panel rotation alleviates this limitation.
For a fixed boresight $\mathbf f(t)$, rotation about
the boresight preserves directional gains while
continuously varying the projected separation over
$[-2\ell\sin a\cos t,\,2\ell\sin a\cos t]$.
Thus, an installed pair achieves zero correlation
whenever $\ell\cos t\geq\ell_{\mathrm{crit}}$,
without antenna movement.
Under unrestricted rotation, a pair of length $\ell$
achieves $\mathcal T(\ell)$.
Since $\mathcal T(\ell)$ is nondecreasing,
a longest installed pair is optimal.
Let $\ell_a=\max_{m<n}\|\mathbf t_m^0-\mathbf t_n^0\|_2$.
If the reference pair belongs to $\mathcal T_0$, then we have
\[
\tau_{\mathrm{rot}}^\star
\leq\tau_{\mathrm{sel+rot}}^\star
=\mathcal T(\ell_a)
\leq\tau_{\mathrm{mov+rot}}^\star.
\]
By Proposition~\ref{prop:III-B-aperture-limits},
discrete activation attains the continuous
movement-rotation benchmark iff
$\ell_a=\ell_{\max}$ or
$\ell_a\geq\ell_{\mathrm{full}}$.

\vspace{-4mm}
\subsection{Multi-AP Channel Structure and Antenna Allocation}
\label{subsec:multi_ap_separation}
\vspace{-2mm}

We extend the LoS analysis to coherent multi-AP
transmission, focusing on distributed channel separation
and antenna allocation.
We consider coherent LoS transmission from $B$ APs
to $K$ users at locations $\{\mathbf q_k\}_{k=1}^{K}$.
Let $\mathcal S_b=\{n:a_{b,n}=1\}$,
$n_b=|\mathcal S_b|$, and
$
\eta_{bk}\triangleq
\beta_{b,0}(\mathbf q_k)
G\!\left(\varepsilon_{k,b,0}\right)$.
The global effective channel of user $k$ is
$
\mathbf g_k
=\operatorname{col}\bigl(
\mathbf g_1(\mathbf q_k),\ldots,
\mathbf g_B(\mathbf q_k)
\bigr)$,
with strength
$s_k=\|\mathbf g_k\|_2^2
=\sum_{b=1}^{B}n_b\eta_{bk}$, where \(\operatorname{col}(\cdot)\) denotes vertical vector concatenation.
The dependence on $\boldsymbol\Theta$ is omitted
for brevity.
For users $i$ and $j$, we define
$\Delta d_{b,ij}=d_b(\mathbf q_i)-d_b(\mathbf q_j)$,
$\Delta\boldsymbol\rho_{b,ij}
=\boldsymbol\rho_{b,0}(\mathbf q_i)
-\boldsymbol\rho_{b,0}(\mathbf q_j)$,
and $\kappa=2\pi/\lambda$.
The local correlation kernel is
\begin{equation}
	F_{b,ij}
	=\frac{1}{n_b}
	\sum\nolimits_{n\in\mathcal S_b}
	e^{\mathrm j\kappa
		\Delta\boldsymbol\rho_{b,ij}^{\mathrm T}
		\mathbf R_b\mathbf t_{b,n}},
	\label{eq:c_local_correlation}
\end{equation}
for $n_b>0$, and $F_{b,ij}=0$ otherwise.
Let
$
z_{b,ij}
=n_b\sqrt{\eta_{bi}\eta_{bj}}\,
e^{-\mathrm j\kappa\Delta d_{b,ij}}F_{b,ij}$.
With
$\mathbf H_b=
[\mathbf g_b(\mathbf q_1),\ldots,
\mathbf g_b(\mathbf q_K)]^{\mathrm T}$,
the local and global Gram matrices are
$\mathbf G_b=\mathbf H_b\mathbf H_b^{\mathrm H}$ and $
\mathbf G=\sum\nolimits_{b=1}^{B}\mathbf G_b$,
where $[\mathbf G_b]_{kk}=n_b\eta_{bk}$ and
$[\mathbf G_b]_{ij}=z_{b,ij}$ for $i\neq j$.
For nonzero global channels, the normalized
correlation between users $i$ and $j$ is
\begin{equation}
	\chi_{ij}
	=\frac{[\mathbf G]_{ij}}{\sqrt{s_i s_j}}
	=\frac{\sum_{b=1}^{B}z_{b,ij}}{\sqrt{s_i s_j}}.
	\label{eq:c_global_correlation}
\end{equation}
The global channels are mutually orthogonal
if and only if
\begin{equation}
	\sum\nolimits_{b=1}^{B}z_{b,ij}=0,
	1\leq i<j\leq K.
	\label{eq:c_global_orthogonality}
\end{equation}
Local orthogonality at every AP is sufficient
but unnecessary. Cross-AP contributions can
cancel even when local correlations are nonzero.
Moreover, complementary AP channel subspaces
can yield a full-rank global Gram matrix even
when individual local Gram matrices are
rank deficient.

To characterize AP-wise antenna allocation,
let $\mathbf H_b^{\mathrm{all}}
\in\mathbb C^{K\times\bar N_b}$ denote
the LoS channel matrix of all installed
antennas at AP $b$, and define
$\overline{\mathbf H}_b
=\mathbf H_b^{\mathrm{all}}/\sqrt{\bar N_b}$.
The normalized reference Gram matrix is
$\mathbf A_b=\overline{\mathbf H}_b
\overline{\mathbf H}_b^{\mathrm H}$.
For a fixed user realization and panel orientation,
uniformly random activation of $n_b$ antennas
yields the conditional mean Gram matrix
$n_b\mathbf A_b$.
Motivated by this relation, we adopt the
fixed-structure continuous-count model
$\mathbf G_b(n_b)=n_b\mathbf A_b$,
where $\mathbf A_b$ remains fixed while
$n_b\geq0$ varies subject to
$\sum_b n_b=N$.
We assume nonbinding per-AP antenna-count limits.
This model preserves the reference spatial
correlation structure while scaling channel
strength with the antenna count.
The extension to $n_b=0$
means that AP $b$ is inactive. Define
\begin{equation}
 \mathbf D=\operatorname{diag}(\sigma_1^2,\ldots,\sigma_K^2),
  \mathbf M=\sum\nolimits_{b=1}^{B}P_b\mathbf A_b,
 \label{eq:c_multiuser_M_T}
\end{equation}
where $P_b>0$ and $\sigma_k^2>0$.
For a positive-semidefinite user Gram matrix
$\mathbf Q=\mathbf H\mathbf H^{\mathrm H}$,
we consider the factorization
$\mathbf H=\mathbf Q^{1/2}\mathbf F$,
where $\mathbf F$ is a partial isometry.
Defining the equivalent precoder
$\mathbf V=\mathbf F\mathbf W$, we have
$\mathbf H\mathbf W=\mathbf Q^{1/2}\mathbf V$.
This transformation preserves the optimal
precoding performance under a sum-power
constraint. Hence, the optimal max-min SINR
under unit sum power is given by
\begin{equation}
 \mathcal T_K(\mathbf Q,\mathbf D)
 =\max_{\substack{\mathbf V\in\mathbb C^{K\times K}\\
                  \|\mathbf V\|_{\mathrm F}^2\le1}}
 \min_k\frac{|[\mathbf Q^{1/2}\mathbf V]_{kk}|^2}
 {\sum_{j\ne k}|[\mathbf Q^{1/2}\mathbf V]_{kj}|^2+\sigma_k^2},
 \label{eq:c_multiuser_sum_power_value}
\end{equation}
The following proposition characterizes the joint optimum over continuous antenna counts and unrestricted linear precoding. Under ZF precoding, it further derives the optimal max–min SINR and closed-form AP-wise antenna allocation, revealing the role of multiuser spatial separation.

\begin{figure}[t]
	\centering
	\includegraphics[width=0.47\textwidth]
	{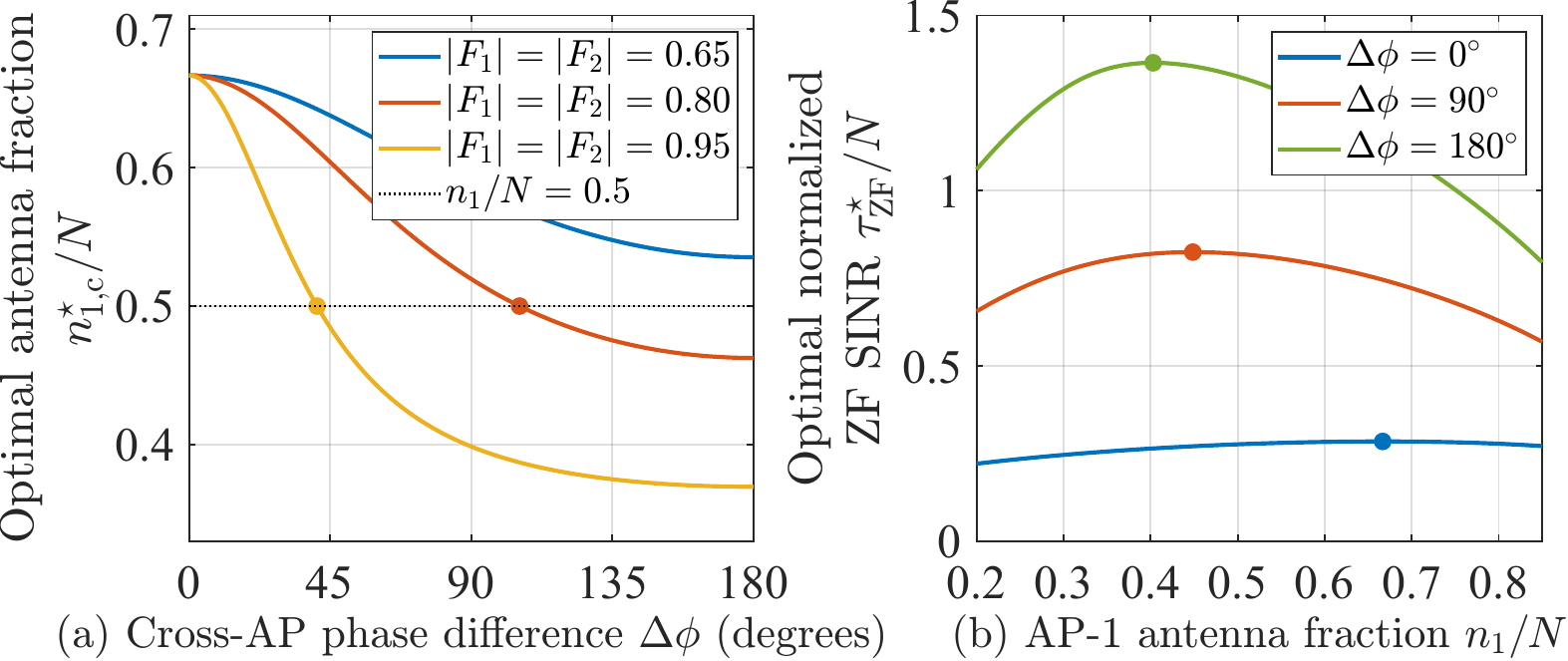}
	\caption{Two-AP, two-user ZF allocation benchmark:
		(a) optimal AP-1 antenna fraction versus the cross-AP
		correlation phase difference; (b) normalized max--min
		SINR versus the AP-1 antenna fraction for $f=0.9$.}
	\label{fig:multi_ap_allocation}
	\vspace{-7mm}
\end{figure}

\begin{proposition}
 \label{prop:c_general_zf_allocation}
 \upshape
 Consider $B$ APs serving $K$ users under the fixed-structure
 continuous-count relaxation and per-AP power budgets
 $\{P_b\}$. Let $\tau_{\mathrm{AP}}^\star(\mathbf n)$ denote
 the optimal max--min SINR with unrestricted linear precoding
 for counts $\mathbf n$. We have
 \begin{equation}
  \tau_{\mathrm{AP,c}}^\star
  \triangleq\max_{{\mathbf n\ge\mathbf0, \sum_b n_b=N}}
   \tau_{\mathrm{AP}}^\star(\mathbf n)
  =\mathcal T_K(N\mathbf M,\mathbf D).
  \label{eq:c_optimal_multiuser_ap_sinr}
 \end{equation}
 If $\mathbf M\succ\mathbf0$, we define
 $T_b=\operatorname{tr}(\mathbf D\mathbf M^{-1}
 \mathbf A_b\mathbf M^{-1})$. Under ZF, the unique
 optimal continuous antenna allocation is
 \begin{equation}
  n_{b,\mathrm c}^\star
  =\frac{NP_bT_b}{\operatorname{tr}(\mathbf D\mathbf M^{-1})},
   b=1,\ldots,B,
  \label{eq:c_optimal_zf_allocation_KB}
 \end{equation}
  where $N$ denotes the total continuous antenna-count
 budget. The resulting optimal minimum-user SINR is
 \begin{equation}
  \tau_{\mathrm{ZF,c}}^\star
  =\frac{N}{\operatorname{tr}(\mathbf D\mathbf M^{-1})}.
  \label{eq:c_general_optimal_zf_sinr}
 \end{equation}
 If $\mathbf M$ is rank deficient, all-user ZF cannot
 achieve a positive minimum SINR for any allocation.
\end{proposition}
\begin{IEEEproof}
 See Appendix~\ref{app:c_zf_allocation}.
\end{IEEEproof}
The equivalence in
\eqref{eq:c_optimal_multiuser_ap_sinr}
shows that the optimal antenna-count fractions
equal the AP-wise power fractions of the
virtual precoder.
For $K=1$, \eqref{eq:c_optimal_zf_allocation_KB}
reduces to
$n_{b,c}^\star
=NP_b\eta_{b1}/\sum_jP_j\eta_{j1}$,
consistent with the continuous relaxation
of the single-user allocation in
Section~\ref{sec:single_user_benchmark}.
For multiple users, the allocation additionally
accounts for spatial separation.
In particular, for equal
user noise powers, let
$\mathbf M=\mathbf U\operatorname{diag}
(\lambda_1,\ldots,\lambda_K)\mathbf U^{\mathrm H}$,
where $\mathbf U=[\mathbf u_1,\ldots,\mathbf u_K]$
is unitary and $\lambda_k>0$. Then, we have
\begin{equation}
	\frac{n_{b,\mathrm c}^\star}{N}
	=\frac{P_b\sum_{k=1}^{K}
		(\mathbf u_k^{\mathrm H}\mathbf A_b\mathbf u_k)/\lambda_k^2}
	{\sum_{k=1}^{K}1/\lambda_k}.
	\label{eq:c_weak_mode_allocation}
\end{equation}
Thus, an AP that strengthens weak spatial modes may receive
more antennas despite a smaller aggregate channel gain.
Antenna activation and panel rotation affect both gains and
correlations through $\mathbf A_b$. Since discrete activation
also changes $\mathbf A_b$ as $n_b$ varies, this allocation
is a fixed-structure benchmark for the original design.
For $K=2$, \eqref{eq:c_general_optimal_zf_sinr} becomes
\begin{equation}
 \tau_{\mathrm{ZF,c}}^\star
 =\frac{N\det\mathbf M}
 {\sigma_1^2M_{22}+\sigma_2^2M_{11}}.
 \label{eq:c_two_user_zf_count_sinr}
\end{equation}
For $B=2$, let $m_b=P_b[\mathbf A_b]_{12}$ and
$\Delta\phi=\arg m_1-\arg m_2$ when both terms are nonzero.
Then
\begin{equation}
 |M_{12}|^2=(|m_1|-|m_2|)^2
 +4|m_1||m_2|\cos^2(\Delta\phi/2).
 \label{eq:c_phase_cancellation}
\end{equation}
With fixed local gains and correlation magnitudes,
increasing $\Delta\phi$ from $0$ to $\pi$ reduces
$|M_{12}|$ and increases $\det\mathbf M$. Hence,
\begin{equation}
 \tau_{\mathrm{ZF,c}}^\star(\pi)
 \ge\tau_{\mathrm{ZF,c}}^\star(\Delta\phi)
 \ge\tau_{\mathrm{ZF,c}}^\star(0).
 \label{eq:c_phase_ordering}
\end{equation}
If $|m_1|=|m_2|$, opposite phases cancel the off-diagonal
term of the virtual-channel Gram matrix. Thus, local
correlations can remain nonzero while their weighted
contributions cancel across APs. This comparison assumes
fixed local gains and correlation magnitudes.
Fig.~\ref{fig:multi_ap_allocation} illustrates
Proposition~\ref{prop:c_general_zf_allocation} for $B=K=2$,
$\eta_{b1}=\eta_{b2}=\eta_b$,
$\sigma_1^2=\sigma_2^2=1$, $|F_1|=|F_2|=f$ with
$F_b=F_{b,12}$,
$P_1\eta_1=2$, and $P_2\eta_2=1$.
Fig.~\ref{fig:multi_ap_allocation}(a) shows that
increasing the phase difference shifts antenna
allocation toward the weaker AP~2.
Fig.~\ref{fig:multi_ap_allocation}(b) confirms that
the SINR peaks occur at the predicted optimal
antenna allocations, while larger phase differences
yield higher optimal ZF SINRs through cross-AP
correlation cancellation.

\vspace{-2mm}
\section{Algorithm Design}
\label{Algorithm Design}
\vspace{-1mm}
In this section, we develop a two-timescale optimization
algorithm based on SAA and alternating optimization.

\vspace{-3mm}
\subsection{Sample-Average Reformulation of the Long-Term Problem}
\label{subsec:long_term_saa}
\vspace{-1mm}
The long-term objective in \eqref{eq:long_term_objective}
averages the optimized minimum instantaneous user rate
over random user sets and channel realizations.
Its evaluation requires averaging over user counts,
locations, and fading realizations, with each realization
involving the short-term precoding optimization in
\eqref{eq:instantaneous_max_min_value}.
Since this expectation is generally intractable to
evaluate analytically, we adopt SAA to obtain a finite-scenario problem~\cite{Shao20256DMA}.
Specifically, we generate $S$ independent nonempty
user-channel scenarios $\{\xi_s\}_{s=1}^{S}$,
each containing the user count $K_s$, user locations,
NLoS departure directions, and small-scale
fading coefficients.
Conditioned on $K_s\geq 1$, the user count follows
\begin{equation}
	\Pr(K_s=k\mid K_s\geq1)
	=
	\frac{\mu^k}{k!(e^\mu-1)},
	 k=1,2,\ldots.
	\label{eq:saa_conditional_user_count}
\end{equation}
Given $K_s$, user locations are independently drawn from
$f(\mathbf q)$ in  \eqref{eq:user_location_density}.
Let $\mathcal U_s$ denote the corresponding user realization.
For scenario $s$, we denote the effective channels by
$\mathbf H_s(\mathbf a,\boldsymbol\vartheta)$ and the
short-term precoders by
$\mathbf W_s=\{\mathbf w_{s,b,j}\}_{b,j}$.
Each scenario has its own precoders, whereas
$\boldsymbol\Theta$ is shared across scenarios. We define
$V_s(\mathbf a,\boldsymbol\vartheta)
	\triangleq
	V\bigl(
	\mathcal U_s,
	\mathbf H_s(\mathbf a,\boldsymbol\vartheta);
	\boldsymbol\Theta
	\bigr)$.
The sample-average objective is
$\widehat F_S(\mathbf a,\boldsymbol\vartheta)
	\triangleq
	\frac{1}{S}\sum\nolimits_{s=1}^{S}
	V_s(\mathbf a,\boldsymbol\vartheta)$.

We introduce a minimum-rate variable
$t_s$ for each scenario.
The sample-average problem is equivalently expressed as
\begin{subequations}
	\label{eq:saa_joint_problem}
	\begin{align}
		\max_{{
				\mathbf a,\boldsymbol\vartheta,
				\{\mathbf W_s,t_s\}_{s=1}^{S}
		}}
		\quad&
		\frac{1}{S}\sum\nolimits_{s=1}^{S}t_s
		\label{eq:saa_joint_objective}
		\\
		\mathrm{s.t.}\quad&
		t_s\leq
		R_{s,k}\bigl(
		\mathbf H_s(\mathbf a,\boldsymbol\vartheta),
		\mathbf W_s
		\bigr),
		 \forall s,k,
		\label{eq:saa_joint_rate}
		\\
		&
		\sum\nolimits_{j=1}^{K_s}
		\|\mathbf w_{s,b,j}\|_2^2
		\leq P_b,
		 \forall s,b,
		\label{eq:saa_joint_power}
		\\
		&
		\mathbf w_{s,b,j}
		=
		\mathbf D_b\mathbf w_{s,b,j},
		 \forall s,b,j,
		\label{eq:saa_joint_mask}\\
		&
		\boldsymbol\vartheta\in\mathcal V, \mathbf a \in 	\mathcal A_{\mathrm{bin}},
		\label{eq:saa_joint_orientation}
	\end{align}
\end{subequations}
where 
\begin{equation}
\!\!\!\!\!	\mathcal A_{\mathrm{bin}}
	=
	\left\{\mathbf a\ \middle|\
	\begin{aligned}
		&a_{b,n}\in\{0,1\}, \sum\nolimits_{n=1}^{\bar N_b}a_{b,n}\leq N_b^{\mathrm{RF}},
		\forall b,\\
		&\sum\nolimits_{b=1}^{B}\sum\nolimits_{n=1}^{\bar N_b}a_{b,n}=N_{\mathrm{act}}
	\end{aligned}\right\}.
	\label{eq:ivd_binary_set}
\end{equation}
For any fixed $(\mathbf a,\boldsymbol\vartheta)$, the
precoding variables and rate constraints are independent
across scenarios. Optimizing $(\mathbf W_s,t_s)$
therefore yields
$t_s^\star=V_s(\mathbf a,\boldsymbol\vartheta)$.
Consequently, eliminating the short-term variables reduces
\eqref{eq:saa_joint_problem} to
$\max_{\mathbf a\in\mathcal A_{\mathrm{bin}},
	\boldsymbol\vartheta\in\mathcal V}
\widehat F_S(\mathbf a,\boldsymbol\vartheta)$.
To solve the resulting problem, we consider three
coupled subproblems involving short-term precoding,
panel orientation, and antenna activation.

\vspace{-4mm}
\subsection{Short-Term Max--Min Beamforming}
\label{subsec:short_term_beamforming}
\vspace{-1mm}
\begingroup

For fixed $(\mathbf a,\boldsymbol\vartheta)$,
the scenarios are independent and each short-term value is
$V=\log_2(1+\tau^\star)$, where $\tau^\star$ is the
optimal minimum user SINR. We solve these problems by
SINR bisection and generalized UL-DL duality under
per-AP power constraints~\cite{YuLan2007,CaiEtAl2012}.
Omitting the scenario index, let
$\mathbf w_k\in\mathbb C^N$, $N=N_{\mathrm{act}}$,
collect the active-antenna precoders, and define
$\mathbf h_k$ such that
$\mathbf h_k^{\mathrm H}\mathbf w_j
=\sum_b\mathbf g_{k,b}^{\mathrm T}\mathbf w_{b,j}$.
Let $\mathbf E_b$ be the diagonal projector onto
AP $b$'s active antennas, with $\sum_b\mathbf E_b=\mathbf I_N$.
For a target SINR $\tau>0$, we consider the power-scaling problem
\begin{align}
 \!\!\!\!   t^\star(\tau)\!=\! \min_{\{\mathbf w_k\},t} &t \nonumber \\
  \!\!\!\!  \!\!\mathrm{s.t.}
 &\gamma_k(\mathbf W)\geq\tau,
 \sum\nolimits_k \!\! \mathbf w_k^{\mathrm H}\mathbf E_b\mathbf w_k
 \leq tP_b,\forall k,\forall b.
\label{eq:short_term_power_scaling}
\end{align}
The target is feasible if and only if
$t^\star(\tau)\leq1$. Phase normalization
yields an equivalent convex problem.
The AP power multipliers satisfy
$\sum_b\nu_bP_b=1$, and
$\lambda_b=\nu_bP_b$ defines normalized
weights on $\Delta_B$.
For a finite optimum, strong duality gives

\begin{equation}
	t^\star(\tau)
	=\max_{\boldsymbol\lambda\in\Delta_B}
	{\min_{\mathbf W:\,\gamma_k(\mathbf W)\geq\tau,\,\forall k}
		\sum\nolimits_k\mathbf w_k^{\mathrm H}
		\mathbf Q(\boldsymbol\lambda)\mathbf w_k},
	\label{eq:short_term_dual_problem}
\end{equation}
where $\Delta_B=\{\boldsymbol\lambda\geq\mathbf0:
\mathbf1^{\mathrm T}\boldsymbol\lambda=1\}$ and
$\mathbf Q(\boldsymbol\lambda)=\sum_b(\lambda_b/P_b)\mathbf E_b$.
Numerically, we use
$\Delta_B^\delta=\{\boldsymbol\lambda\in\Delta_B:
\lambda_b\geq\delta,\ \forall b\}$, $0<\delta<1/B$,
to ensure $\mathbf Q\succ\mathbf0$, reducing $\delta$
when tighter bounds are needed.

For fixed $\boldsymbol\lambda$, let $q_k$ denote the
virtual uplink power and define
$\overline{\mathbf h}_k=\mathbf h_k/\sigma_k$,
$\overline{\mathbf H}=[\overline{\mathbf h}_1,\ldots,
\overline{\mathbf h}_K]$,
$\mathbf D_q=\operatorname{diag}(q_1,\ldots,q_K)$,
and $\mathbf G=\overline{\mathbf H}^{\mathrm H}
\mathbf Q^{-1}\overline{\mathbf H}$.
For a feasible inner problem with finite optimum,
the uplink SINR equations give $q_k=\tau/d_k$, where
\[
 d_k=\overline{\mathbf h}_k^{\mathrm H}
 \left(\mathbf Q+\sum\nolimits_{j\neq k}q_j
 \overline{\mathbf h}_j\overline{\mathbf h}_j^{\mathrm H}\right)^{-1}
 \overline{\mathbf h}_k.
\]
The Woodbury and Sherman-Morrison identities yield
$[\mathbf T]_{kk}=d_k/(1+q_kd_k)$, giving the
fixed-point iteration initialized with $q_k^{(0)}>0$,
\begin{align}
 \mathbf T^{(i)}
 &=\mathbf G(\mathbf I_K+\mathbf D_q^{(i)}\mathbf G)^{-1},
 q_k^{(i+1)}
 =\tau\left(\frac{1}{[\mathbf T^{(i)}]_{kk}}-q_k^{(i)}\right). \nonumber
\end{align}
Thus, each iteration solves a $K\times K$ linear system,
with $\mathbf G$ reused within each dual update.
At the converged powers, the MMSE directions are the
normalized columns of
$\mathbf Q^{-1}\overline{\mathbf H}
(\mathbf I_K+\mathbf D_q\mathbf G)^{-1}$.
Let $\mathbf v_k$ denote these directions and define
$[\mathbf F]_{kk}=0$,
$[\mathbf F]_{kj}=|\overline{\mathbf h}_k^{\mathrm H}\mathbf v_j|^2/
|\overline{\mathbf h}_k^{\mathrm H}\mathbf v_k|^2$ for $j\neq k$,
and $u_k=|\overline{\mathbf h}_k^{\mathrm H}\mathbf v_k|^{-2}$.
Enforcing the target SINRs gives
$\mathbf p=\tau\mathbf F\mathbf p+\tau\mathbf u$,
and hence the downlink precoders are recovered as
\begin{equation}
 \mathbf p=\tau(\mathbf I_K-\tau\mathbf F)^{-1}\mathbf u,
  \mathbf w_k=\sqrt{p_k}\mathbf v_k,
 \label{eq:short_term_dl_power}
\end{equation}
provided $\rho(\tau\mathbf F)<1$. Here, $\mathbf p=[p_1,\ldots,p_K]^T$ and
$\mathbf u=[u_1,\ldots,u_K]^T$, where $p_k$
denotes the downlink power allocated to user $k$.
The notation $\rho(\cdot)$ denotes the spectral
radius of a matrix.
For an exact inner solution,
$r_b=P_b^{-1}\sum_k\mathbf w_k^{\mathrm H}
\mathbf E_b\mathbf w_k$
is the $b$-th component of a supergradient
of the dual objective in \eqref{eq:short_term_dual_problem}.
The AP weights are updated by
\begin{equation}
 \boldsymbol\lambda^{(i+1)}
 =\Pi_{\Delta_B^\delta}
 \bigl(\boldsymbol\lambda^{(i)}+\eta_i\mathbf r^{(i)}\bigr),
 \label{eq:short_term_dual_update}
\end{equation}
with $\eta_i>0$, $\sum_i\eta_i=\infty$, and
$\sum_i\eta_i^2<\infty$. Here, $\Pi_{\Delta_B^\delta}(\cdot)$ denotes
the Euclidean projection onto $\Delta_B^\delta$.
At each update, we solve the inner problem to the required
accuracy.

To determine whether $\tau$ is feasible, we bound
$t^\star(\tau)$ using downlink precoders that satisfy
all target SINRs and nonnegative uplink powers
$\mathbf q$ that satisfy
\begin{equation}
 \mathbf Q+\sum\nolimits_{j\neq k}q_j
 \overline{\mathbf h}_j\overline{\mathbf h}_j^{\mathrm H}
 -\frac{q_k}{\tau}
 \overline{\mathbf h}_k\overline{\mathbf h}_k^{\mathrm H}
 \succeq\mathbf0,\forall k,
 \label{eq:short_term_dual_feasibility}
\end{equation}
These primal- and dual-feasible solutions yield
\begin{equation}
 \underline t(\tau)\triangleq\sum\nolimits_k q_k
 \leq t^\star(\tau)
 \leq\max_b r_b\triangleq\overline t(\tau).
 \label{eq:short_term_scaling_bounds}
\end{equation}
We check the downlink SINRs and dual constraints
before evaluating the bounds.
If the dual constraints are violated, we scale down
$\mathbf q$ to compute a valid lower bound.
The upper bound uses the recovered downlink precoders.
The target is feasible if $\overline t\leq1$ and
infeasible if $\underline t>1$.
Otherwise, we continue the iterations and reduce
$\delta$ as needed.
Bisection starts from $\tau_{\mathrm L}=0$ and
\begin{equation}
 \tau_{\mathrm U}
 =\min_k\frac{\bigl(\sum_b\sqrt{P_b}
 \|\mathbf g_{k,b}\|_2\bigr)^2}{\sigma_k^2}.
 \label{eq:short_term_sinr_upper_bound}
\end{equation}
If $\tau_{\mathrm U}=0$, then $\tau^\star=0$.
Otherwise, we test the midpoint
$\tau=(\tau_{\mathrm L}+\tau_{\mathrm U})/2$.
If feasible, we set $\tau_{\mathrm L}=\tau$
and retain the corresponding precoders.
If infeasible, we set $\tau_{\mathrm U}=\tau$.
We stop when
$\tau_{\mathrm U}-\tau_{\mathrm L}\leq\epsilon_{\mathrm{SINR}}$
or the iteration limit is reached.
The current bounds are retained if feasibility
remains undecided.
Using the SINR bounds
$\tau_{s,\mathrm L}(\boldsymbol\Theta)$ and
$\tau_{s,\mathrm U}(\boldsymbol\Theta)$
for each scenario, we define
\begin{align}
 F_L(\boldsymbol\Theta)
 &=\frac1S\sum\nolimits_{s=1}^{S}
 \log_2\bigl(1+\tau_{s,\mathrm L}(\boldsymbol\Theta)\bigr),
 \label{eq:saa_objective_bounds1}\\
 F_U(\boldsymbol\Theta)
 &=\frac1S\sum\nolimits_{s=1}^{S}
 \log_2\bigl(1+\tau_{s,\mathrm U}(\boldsymbol\Theta)\bigr).
 \label{eq:saa_objective_bounds}
\end{align}
{The bounds $F_L\leq\widehat F_S\leq F_U$
certify long-term updates.}
\par
\endgroup

\vspace{-2mm}

\begingroup
\vspace{-2mm}
\subsection{Panel-Orientation Optimization}
\label{subsec:panel_orientation_optimization}
\vspace{-1mm}
For fixed $\mathbf a$, we optimize
$\widehat F_S(\mathbf a,\boldsymbol\vartheta)$ over
$\boldsymbol\vartheta\in\mathcal V$.
Starting from the current feasible orientation,
we solve the short-term precoding problems at
$\boldsymbol\vartheta^{(i)}$ using
Section~\ref{subsec:short_term_beamforming}.
We fix the resulting feasible precoders
$\mathbf W_s^{(i)}$ and construct a WMMSE-based rate lower bound.
We define the received coefficients
\begin{equation}
 c_{s,k,j}^{(i)}(\boldsymbol\vartheta)
 =\sum\nolimits_{b=1}^{B}
 \mathbf g_{s,k,b}(\mathbf a,\boldsymbol\vartheta)^{\mathrm T}
 \mathbf w_{s,b,j}^{(i)}.
 \label{eq:orientation_received_coefficient}
\end{equation}
For a receiver coefficient $u$, the mean-square error is
\begin{equation}
\begin{aligned}
 e_{s,k}^{(i)}(\boldsymbol\vartheta,u)
 ={}&|u|^2\left(\sum\nolimits_{j=1}^{K_s}
 |c_{s,k,j}^{(i)}(\boldsymbol\vartheta)|^2+\sigma_{s,k}^2\right)\\
 &-2\operatorname{Re}\{u^*c_{s,k,k}^{(i)}(\boldsymbol\vartheta)\}+1.
\end{aligned}
\label{eq:orientation_user_mse}
\end{equation}
At $\boldsymbol\vartheta^{(i)}$, we set
$T_{s,k}^{(i)}=\sum_j|c_{s,k,j}^{(i)}
(\boldsymbol\vartheta^{(i)})|^2+\sigma_{s,k}^2$ and update
\begin{equation}
 u_{s,k}^{(i)}
 =\frac{c_{s,k,k}^{(i)}(\boldsymbol\vartheta^{(i)})}{T_{s,k}^{(i)}},
 v_{s,k}^{(i)}
 =\frac{1}{e_{s,k}^{(i)}(\boldsymbol\vartheta^{(i)},u_{s,k}^{(i)})}.
 \label{eq:orientation_mmse_auxiliary_update}
\end{equation}
With these variables fixed, the WMMSE identity gives
\begin{equation}
\begin{aligned}
 \phi_{s,k}^{(i)}(\boldsymbol\vartheta)
 &=\frac{\ln v_{s,k}^{(i)}
 -v_{s,k}^{(i)}e_{s,k}^{(i)}(\boldsymbol\vartheta,u_{s,k}^{(i)})+1}{\ln2}\\
 &\leq R_{s,k}\bigl(\mathbf H_s(\mathbf a,\boldsymbol\vartheta),
 \mathbf W_s^{(i)}\bigr).
\end{aligned}
\label{eq:orientation_wmmse_rate_lower_bound}
\end{equation}
Thus, $\min_k\phi_{s,k}^{(i)}(\boldsymbol\vartheta)
\leq V_s(\mathbf a,\boldsymbol\vartheta)$,
with equality at $\boldsymbol\vartheta^{(i)}$
for exact short-term precoders.
The surrogate gradient is
$\mathbf d_{s,k}^{(i)}
=\nabla_{\boldsymbol\vartheta}\phi_{s,k}^{(i)}
(\boldsymbol\vartheta^{(i)})
=-(v_{s,k}^{(i)}/\ln2)\nabla_{\boldsymbol\vartheta}
e_{s,k}^{(i)}(\boldsymbol\vartheta^{(i)},u_{s,k}^{(i)})$.
For $\xi_b\in\{\alpha_b,\beta_b,\gamma_b\}$, we obtain
\begin{align}
 \! \frac{\partial e_{s,k}^{(i)}}{\partial\xi_b}
 \!=\! 2\operatorname{Re} \! \Bigg\{\!
 |u_{s,k}^{(i)}|^2\sum\nolimits_{j=1}^{K_s} \!
 (c_{s,k,j}^{(i)})^*
 \frac{\partial c_{s,k,j}^{(i)}}{\partial\xi_b}
 \!-\!\! (u_{s,k}^{(i)})^*
 \frac{\partial c_{s,k,k}^{(i)}}{\partial\xi_b}\! \Bigg\},
\nonumber
\end{align}
where the angle arguments are omitted and
\begin{equation}
 \frac{\partial c_{s,k,j}^{(i)}}{\partial\xi_b}
 =\left(\frac{\partial\mathbf g_{s,k,b}}{\partial\xi_b}\right)^{\mathrm T}
 \mathbf w_{s,b,j}^{(i)}.
 \label{eq:orientation_coefficient_derivative}
\end{equation}
The channel derivatives in \eqref{eq:orientation_coefficient_derivative} follow directly
from the chain rule applied to the rotation
matrix in \eqref{eq:panel_rotation_parameterization} and the channel model in
\eqref{eq:full_array_response}-\eqref{eq:full_ap_user_channel}, and are omitted for brevity.


Let $\Delta\boldsymbol\vartheta
=\boldsymbol\vartheta-\boldsymbol\vartheta^{(i)}$.
We use the concave quadratic surrogate
\begin{equation}
\begin{aligned}
\!\!\! q_{s,k}^{(i)}(\boldsymbol\vartheta;L_{s,k})
 \!=&\phi_{s,k}^{(i)}(\boldsymbol\vartheta^{(i)})
 \!+\!(\mathbf d_{s,k}^{(i)})^{\mathrm T}\Delta\boldsymbol\vartheta
 \!-\!\frac{L_{s,k}}2\|\Delta\boldsymbol\vartheta\|_2^2,
\end{aligned}
\label{eq:orientation_quadratic_surrogate}
\end{equation}
with $L_{s,k}>0$ chosen by backtracking.
The closed convex search region $\mathcal V^{(i)}$
contains $\boldsymbol\vartheta^{(i)}$, and the gradients
of $\phi_{s,k}^{(i)}$ are Lipschitz continuous on this region.
For the adopted directional field gain, we use
$\mathcal V^{(i)}=\mathcal V$ when $p\geq2$.
For $0<p<2$, we use a local angle box within
$\mathcal V$ that avoids nonsmooth pattern boundaries.
The trial orientation solves
\begin{equation}
\begin{aligned}
 \max_{\boldsymbol\vartheta,\{t_s\}}\quad
 &\frac1S\sum\nolimits_{s=1}^{S}t_s\\
 \mathrm{s.t.}\quad
 &t_s\leq q_{s,k}^{(i)}(\boldsymbol\vartheta;L_{s,k}),
 \boldsymbol\vartheta\in\mathcal V^{(i)}, \forall s,k.
\end{aligned}
\label{eq:orientation_convex_update}
\end{equation}
This problem is convex. Its solution
$\widetilde{\boldsymbol\vartheta}$ must satisfy
\begin{equation}
 q_{s,k}^{(i)}(\widetilde{\boldsymbol\vartheta};L_{s,k})
 \leq\phi_{s,k}^{(i)}(\widetilde{\boldsymbol\vartheta}),
 \forall s,k.
 \label{eq:orientation_curvature_test}
\end{equation}
If a condition fails, we increase the corresponding
$L_{s,k}\leftarrow\eta L_{s,k}$, $\eta>1$, and solve
\eqref{eq:orientation_convex_update} again, keeping
the precoders, auxiliary variables, and gradients fixed.
Lipschitz continuity ensures finite backtracking.
We then reoptimize the short-term precoders and
accept the candidate only if
$F_L(\mathbf a,\widetilde{\boldsymbol\vartheta})
>F_U(\mathbf a,\boldsymbol\vartheta^{(i)})
+\epsilon_{\mathrm{acc}}$.
If accepted, we set
$\boldsymbol\vartheta^{(i+1)}=\widetilde{\boldsymbol\vartheta}$
and use the reoptimized precoders to update the
auxiliary variables and gradients for the next iteration.
Otherwise, we retain the current orientation and end
this orientation update.
We also stop when the prescribed iteration limit is reached.
\par
\endgroup

\vspace{-4mm}
\subsection{Antenna Activation and Allocation}
\label{subsec:ivd_activation}
\vspace{-1mm}
For fixed panel orientations $\boldsymbol\vartheta$,
the antenna-activation subproblem is formulated as
	\begin{align}
		\max_{\mathbf a}\quad&
	\widehat F_S(\mathbf a,\boldsymbol\vartheta) &
	\mathrm{s.t.} \quad
	\mathbf a\in\mathcal A_{\mathrm{bin}}.
	\label{eq:ivd_binary_problem}
	\end{align}
The scenario-specific values
$V_s(\mathbf a,\boldsymbol\vartheta)$
account for optimal short-term precoding.
We solve the combinatorial subproblem
using a sampled multi-start greedy algorithm.
For initialization, we first relax the activation
variables to $[0,1]$ and couple them with the
precoders through per-antenna power constraints.
The relaxation is solved once at the initial
panel orientations using WMMSE-based successive
convex approximation. Rounding and random
perturbations yield feasible activation patterns.
Motivated by Proposition~\ref{prop:c_general_zf_allocation},
we also construct a ZF-guided initialization
by averaging the analytical antenna-allocation
ratios over sampled scenarios and projecting
them onto feasible integer AP-wise counts.
The corresponding active positions are selected
according to the relaxed activation values,
yielding $\mathbf a^{\mathrm{ZF}}$.
After removing duplicates, the reusable seed set is
\begin{equation}
	\mathcal I_{\mathrm{seed}}
	=
	\{\mathbf a^{\mathrm{rnd}},\mathbf a^{\mathrm{ZF}}\}
	\cup\mathcal I_{\mathrm{rand}},
	|\mathcal I_{\mathrm{seed}}|
	\leq S_{\max}-1.
\end{equation}
Here, $\mathbf a^{\mathrm{rnd}}$ is the rounded
activation-relaxation solution,
$\mathcal I_{\mathrm{rand}}$ contains feasible
randomly perturbed patterns, and $S_{\max}$
is the maximum number of activation-search starts.
At each activation update, we construct
the multi-start set
\begin{equation}
	\mathcal I
	=
	\{\mathbf a^{\mathrm{base}}\}
	\cup\mathcal I_{\mathrm{seed}},
	|\mathcal I|\leq S_{\max},
	\label{eq:activation_multistart}
\end{equation}
where $\mathbf a^{\mathrm{base}}$ is the current activation pattern and
all starts are evaluated under the current panel orientations.
For the $\ell$-th start, let $\mathbf a^{(r)}$
denote the current binary pattern at iteration
$r$. We define the active and inactive position
sets of AP $b$ as
\begin{equation}
	\begin{aligned}
		\mathcal S_b^{(r)}
		&=\{n:a_{b,n}^{(r)}=1\},
		\mathcal S_{b,0}^{(r)}
		=\{n:a_{b,n}^{(r)}=0\}.
	\end{aligned}
	\label{eq:activation_position_sets}
\end{equation}
Let $\mathbf e_{b,n}$ denote the unit vector
related to position $(b,n)$.
We generate candidate patterns through two
local moves.

\emph{1) Intra-AP replacement:}
We replace one active position of AP $b$ with
an inactive position at the same AP:
\begin{equation}
	\widetilde{\mathbf a}
	=
	\mathbf a^{(r)}
	-\mathbf e_{b,n}
	+\mathbf e_{b,m},
	\label{eq:activation_replacement}
\end{equation}
where $n\in\mathcal S_b^{(r)}$ and
$m\in\mathcal S_{b,0}^{(r)}$.

\emph{2) Inter-AP reallocation:}
We transfer one active antenna from AP $b$
to an inactive position of AP $j$:
\begin{equation}
	\widetilde{\mathbf a}
	=
	\mathbf a^{(r)}
	-\mathbf e_{b,n}
	+\mathbf e_{j,m},
	\label{eq:activation_reallocation}
\end{equation}
where $b\ne j$, $n\in\mathcal S_b^{(r)}$,
$m\in\mathcal S_{j,0}^{(r)}$, and
$n_j^{(r)}<N_j^{\mathrm{RF}}$.
This move adjusts the AP-wise antenna counts.
Inter-AP reallocations match AP-wise counts, and subsequent
intra-AP replacements match positions, connecting any two
patterns in $\mathcal A_{\mathrm{bin}}$.

Let $\mathcal N_{\mathrm{rep}}(\mathbf a^{(r)})$
and $\mathcal N_{\mathrm{rea}}(\mathbf a^{(r)})$
denote the feasible neighborhoods generated by
intra-AP replacement and inter-AP reallocation,
respectively, where $r$ indexes accepted
activation updates.
For each incumbent $\mathbf a^{(r)}$,
we examine up to $Q_{\max}$ batches indexed
by $q$, each containing up to $L_{\mathrm{rep}}$
and $L_{\mathrm{rea}}$ candidates uniformly
sampled without replacement from the
respective neighborhoods.
The sampled candidate set is
\begin{equation}
	\!\!\!	\mathcal N_L^{(r,q)}
	\!	=\!
	\mathcal N_{\mathrm{rep}}^{(r,q)}
	\!	\cup\!
	\mathcal N_{\mathrm{rea}}^{(r,q)},
	|\mathcal N_L^{(r,q)}|
	\!	\leq \!
	L_{\mathrm{rep}}+L_{\mathrm{rea}}
	\!	\triangleq \!L_{\mathrm{neigh}}.
\end{equation}

For each candidate $\widetilde{\mathbf a}$,
we reoptimize the short-term precoders
using Section~\ref{subsec:short_term_beamforming}
and compute the SAA bounds in
\eqref{eq:saa_objective_bounds1}
and \eqref{eq:saa_objective_bounds}.
The best candidate in batch $q$ is
\begin{equation}
	\widehat{\mathbf a}^{(r,q)}
	\in
	\arg\max_{\widetilde{\mathbf a}
		\in\mathcal N_L^{(r,q)}}
	F_L(\widetilde{\mathbf a},
	\boldsymbol\vartheta).
	\label{eq:activation_best_candidate}
\end{equation}
The candidate is accepted only if
\begin{equation}
	F_L(\widehat{\mathbf a}^{(r,q)},
	\boldsymbol\vartheta)
	>
	F_U(\mathbf a^{(r)},\boldsymbol\vartheta)
	+\epsilon_{\mathrm{acc}}.
	\label{eq:activation_acceptance}
\end{equation}
If accepted, we set
$\mathbf a^{(r+1)}=\widehat{\mathbf a}^{(r,q)}$
and restart the neighborhood search.
Otherwise, we examine the next batch,
up to $Q_{\max}$ batches per incumbent.
Each local search terminates upon reaching
$R_{\max}$ accepted updates or exhausting
$Q_{\max}$ consecutive batches without
an accepted candidate.
Starting from $\mathbf a^{\mathrm{base}}$,
we compare the terminal patterns from all
starts and retain only certified improvements
under \eqref{eq:activation_acceptance}.
The resulting pattern $\mathbf a^+$ satisfies
$
\widehat F_S(\mathbf a^+,\boldsymbol\vartheta)
\geq
\widehat F_S(\mathbf a^{\mathrm{base}},
\boldsymbol\vartheta).
$

\vspace{-4mm}
\subsection{Computational Complexity and Convergence}
\label{subsec:complexity_convergence}
\vspace{-1mm}

The overall two-timescale workflow is summarized in Algorithm
\ref{alg:joint_panel_activation}.
We analyze the computational complexity and convergence of
the proposed algorithm.

\begin{algorithm}[t]
	\caption{Two-Timescale Optimization Algorithm}
	\label{alg:joint_panel_activation}
	\begin{algorithmic}[1]
		\Require User and channel statistics, sample size $S$,
		feasible initial configuration
		$(\mathbf a^{(0)},\boldsymbol\vartheta^{(0)})$,
		tolerances $\epsilon_{\mathrm{acc}}$ and
		$\epsilon_{\mathrm{out}}$, and iteration budgets.
		
		\State Generate $S$ nonempty
		user-channel scenarios.
		
		\State Generate the initialization set \(\mathcal I_{\mathrm{seed}}\) using activation relaxation, ZF-guided allocation, and random perturbations.
		
		\State Evaluate $\boldsymbol\Theta^{(0)}$
		using short-term precoding optimization
		to obtain $F_L^{(0)}$ and $F_U^{(0)}$ using \eqref{eq:saa_objective_bounds1} and \eqref{eq:saa_objective_bounds}.
		
		\State Set $t\gets0$.
		\Repeat
		\State Update $\boldsymbol\vartheta^{(t+1)}$
		via the WMMSE-based orientation
		optimization in Section~\ref{subsec:panel_orientation_optimization}.
		
		\State Update $\mathbf a^{(t+1)}$
		via the multi-start activation search
		in Section~\ref{subsec:ivd_activation}.
		
		\State Evaluate $\boldsymbol\Theta^{(t+1)}$
		using short-term precoding optimization
		to obtain $F_L^{(t+1)}$ and
		$F_U^{(t+1)}$.
		
		\State Set $t\gets t+1$.
		\Until{$F_U^{(t)}-F_L^{(t-1)}
			\leq\epsilon_{\mathrm{out}}$
			or $t\geq T_{\max}$}		
		\State \Return $\boldsymbol\Theta^\star
		=(\mathbf a^{(t)},\boldsymbol\vartheta^{(t)})$.
		For each channel realization, obtain
		$\mathbf W^\star$ via the
		precoding method in Section~\ref{subsec:short_term_beamforming}.
		
	\end{algorithmic}
\end{algorithm}

\subsubsection{Computational Complexity}
Let $N=N_{\mathrm{act}}$, and let
$J_s$, $I_{\lambda,s}$, and $I_{q,s}$
denote the numbers of SINR bisection,
dual, and uplink fixed-point iterations,
respectively. For scenario $s$,
constructing the Gram matrix requires
$O(NK_s^2)$ operations, while each
fixed-point iteration requires $O(K_s^3)$.
Including downlink precoder reconstruction,
the configuration-evaluation complexity is
\begin{equation}
	\mathcal C_{\mathrm{eval}}
	=\mathcal O\left(
	\sum\nolimits_{s=1}^{S}
	J_s I_{\lambda,s}
	\left[
	NK_s^2+(I_{q,s}+1)K_s^3
	\right]
	\right).
	\label{eq:complexity_evaluation}
\end{equation}
The orientation surrogate has $3B+S$ variables
and $K_\Sigma\triangleq\sum_{s=1}^{S}K_s$ user-rate constraints.
Let $\mathcal C_{\mathrm{ori}}$ denote its
solution cost, $I_{\mathrm{ori}}$ the number
of orientation updates per outer iteration,
and $J_{\mathrm{bt}}$ the maximum number
of backtracking trials.
The activation relaxation is solved once,
with $I_{\mathrm{rel}}$ iterations at a cost
of $\mathcal C_{\mathrm{rel}}$ per iteration.
For the discrete multi-start search,
each of the at most $S_{\max}$ starts
performs up to $R_{\max}$ accepted updates,
with at most $Q_{\max}$ batches of
$L_{\mathrm{neigh}}$ candidates examined
per incumbent.
Thus, the number of candidate evaluations
per activation search is bounded by
$
M_{\mathrm{act}}
=S_{\max}(R_{\max}+1)
Q_{\max}L_{\mathrm{neigh}}$.
The dominant overall complexity is
\begin{equation}
	\begin{aligned}
		\mathcal C_{\mathrm{total}}
		=\mathcal O\Big(
		&I_{\mathrm{rel}}\mathcal C_{\mathrm{rel}}
		+T_{\max}\big[
		I_{\mathrm{ori}}J_{\mathrm{bt}}
		\mathcal C_{\mathrm{ori}}\\
		&+(I_{\mathrm{ori}}+M_{\mathrm{act}}+1)
		\mathcal C_{\mathrm{eval}}
		\big]\Big).
	\end{aligned}
	\label{eq:complexity_total}
\end{equation}

\subsubsection{Convergence Analysis}
Under exact short-term precoding, the WMMSE-based
orientation surrogates are tight at the current
iterate. To ensure monotonicity under finite-accuracy
evaluations, an orientation or activation update
is accepted only if the candidate's certified SAA
lower bound exceeds the incumbent's upper bound
by more than $\epsilon_{\mathrm{acc}}$.
Otherwise, the incumbent is retained.
Thus, the outer iterations satisfy
\begin{equation}
	\widehat F_S(\boldsymbol\Theta^{(t+1)})
	\geq
	\widehat F_S(\mathbf a^{(t)},
	\boldsymbol\vartheta^{(t+1)})
	\geq
	\widehat F_S(\boldsymbol\Theta^{(t)}).
	\label{eq:overall_monotonicity}
\end{equation}
Since the SAA objective is bounded above over
the feasible set, the resulting monotone
sequence converges to a finite limit.
Moreover, the certified objective bounds
imply
\begin{equation}
		0
		\leq
		\widehat F_S(\boldsymbol\Theta^{(t)})
		-\widehat F_S(\boldsymbol\Theta^{(t-1)})
		\leq F_U^{(t)}-F_L^{(t-1)}.
	\label{eq:certified_outer_improvement}
\end{equation}
Thus, the stopping criterion in
Algorithm~\ref{alg:joint_panel_activation} bounds the improvement
between successive iterations by
$\epsilon_{\mathrm{out}}$.

\vspace{-3mm}
\section{Numerical Results }
\label{subsec:simulation_setup}
\vspace{-1mm}

In this section, 
simulation results are provided to verify the effectiveness of the proposed design in a cell-free network. 

\vspace{-4mm}
\subsection{Simulation Setup}
\vspace{-1mm}
In the simulations, 
we consider a downlink cell-free network with
$B=4$ APs uniformly deployed on a circle of
radius $R_{\mathrm A}=140\,\mathrm m$ at a height
of $h=20\,\mathrm m$, serving users within a
ground disk of radius $R_{\mathrm U}=120\,\mathrm m$.
The AP positions are
$\mathbf p_b=
[R_{\mathrm A}\cos\phi_b,
R_{\mathrm A}\sin\phi_b,h]^{\mathrm T}$,
where $\phi_b=2\pi(b-1)/B$.
Users follow a nonhomogeneous PPP with mean
population $\mu=10$ and two circular hotspots
centered at
$\mathbf c_1=[-30,-15,0]^{\mathrm T}\,\mathrm m$
and
$\mathbf c_2=[75,40,0]^{\mathrm T}\,\mathrm m$,
with radii $20\,\mathrm m$ and $15\,\mathrm m$,
respectively.
The spatial intensity is
\begin{equation}
	\rho(\mathbf q)
	=\frac{\xi\mu}{|\mathcal L|}
	+\sum\nolimits_{w=1}^{2}
	\frac{(1-\xi)\mu\omega_w}{|\mathcal L_w|}
	\mathbf 1_{\{\mathbf q\in\mathcal L_w\}},
	\label{eq:sim_user_intensity}
\end{equation}
where $\mathcal L$ is the service disk,
$\mathcal L_w$ denotes hotspot $w$,
$\xi=0.4$ is the background user ratio,
and $(\omega_1,\omega_2)=(2/3,1/3)$.

Each AP is equipped with a $6\times6$ installed
UPA with half-wavelength spacing and at most
$N_b^{\mathrm{RF}}=12$ active antennas.
The default network-wide activation budget
is $N_{\mathrm{act}}=24$.
We set the carrier frequency to
$f_c=28\,\mathrm{GHz}$ and the directional
element parameter to $p=3$.
The reference panel boresight points toward
the service center, with
$\mathbf f_b^{(0)}
=-\mathbf p_b/\|\mathbf p_b\|_2$.
The in-plane reference axes are
$\mathbf u_{b,1}
=[-\sin\phi_b,\cos\phi_b,0]^{\mathrm T}$
and
$\mathbf u_{b,2}
=\mathbf f_b^{(0)}\times\mathbf u_{b,1}$.
The rotation ranges are
$\alpha_b,\beta_b\in[-60^\circ,60^\circ]$
and $\gamma_b\in[-90^\circ,90^\circ]$.
We adopt a far-field multipath channel with
one LoS and $L=3$ NLoS paths per AP-user link.
The path powers follow a Rician factor of
$\kappa_{\mathrm R}=9~\mathrm{dB}$, with equal
NLoS power allocation.
The NLoS departure directions have uniformly
distributed azimuths and truncated half-normal
angular offsets from the LoS direction, with
scale $\sigma_\delta=20^\circ$ and support
$[0,\pi/2]$.
Independent $z_l\sim\mathcal{CN}(0,1)$ model
NLoS fading, and all sampled path parameters
remain fixed during configuration optimization.
The reference distance and path-loss exponent
are set to $d_0=1$~m and $\nu_{\mathrm{PL}}=2.2$,
respectively.
The per-AP transmit power and user
noise power are $P_b=20\,\mathrm{dBm}$
and $\sigma_k^2=-80\,\mathrm{dBm}$,
respectively. 
The SAA sample size is set to $S=500$,
with the sampled user-channel scenarios
fixed throughout the optimization.

We compare the proposed design with four benchmarks:
1) \textbf{Fixed}, with fixed antenna activation
and panel orientations under equal AP-wise
antenna counts;
2) \textbf{Rotation-only}, optimizing panel
orientations with fixed antenna activation;
3) \textbf{Activation-only}, optimizing antenna activation patterns and AP-wise activation counts
with fixed panel orientations; and
4) \textbf{Equal-count joint design}, jointly
optimizing antenna activation and panel
orientations with fixed AP-wise counts
$n_b=N_{\mathrm{act}}/B$.
All schemes employ the same short-term
precoding method and are evaluated
under identical user-channel scenarios.

\vspace{-3mm}
\subsection{Performance Analysis}
\vspace{-1mm}
\begin{figure}[t]
	\centering
	\includegraphics[width=0.42\textwidth]
	{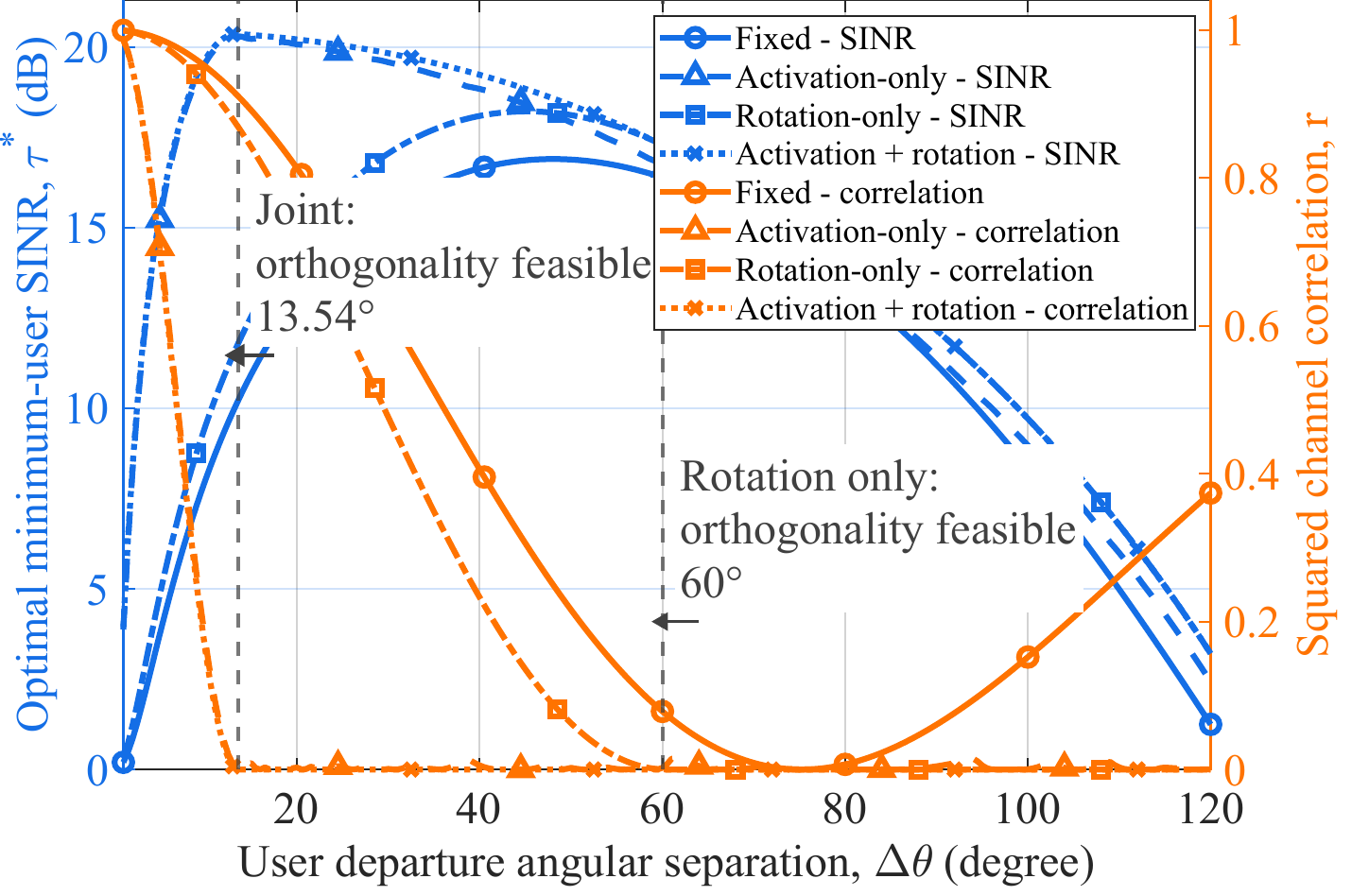}
	\caption{Optimal minimum-user SINR and squared
		channel correlation versus the user departure
		angular separation ($p=3$).}
	\label{fig:angle_p3}
	\vspace{-7mm}
\end{figure}

Fig.~\ref{fig:angle_p3} illustrates the effects of
antenna activation and panel rotation on the
minimum-user SINR and channel correlation
under asymmetric LoS channels ($p=3$).
We consider a single AP equipped with a
$4\times4$ planar array with half-wavelength spacing,
from which two antennas are activated.
Consistent with Proposition~\ref{prop:III-B-aperture-limits},
joint antenna activation and rotation achieve
channel orthogonality at approximately $13.54^\circ$,
compared with $60^\circ$ for rotation alone,
demonstrating the spatial resolution benefit
of activating widely separated antennas.
The activation-only and joint designs achieve
nearly identical performance, as antenna
activation enables near-orthogonal channels
while the fixed panel orientation already
provides nearly optimal directional gains.
Moreover, despite maintaining channel
orthogonality, the joint design's SINR decreases
from $20.22$~dB at $20^\circ$ to $12.17$~dB
at $90^\circ$, due to reduced directional
gains toward widely separated users.

\vspace{-1mm}
\begin{figure}[t]
	\centering
	\includegraphics[width=0.4\textwidth]
	{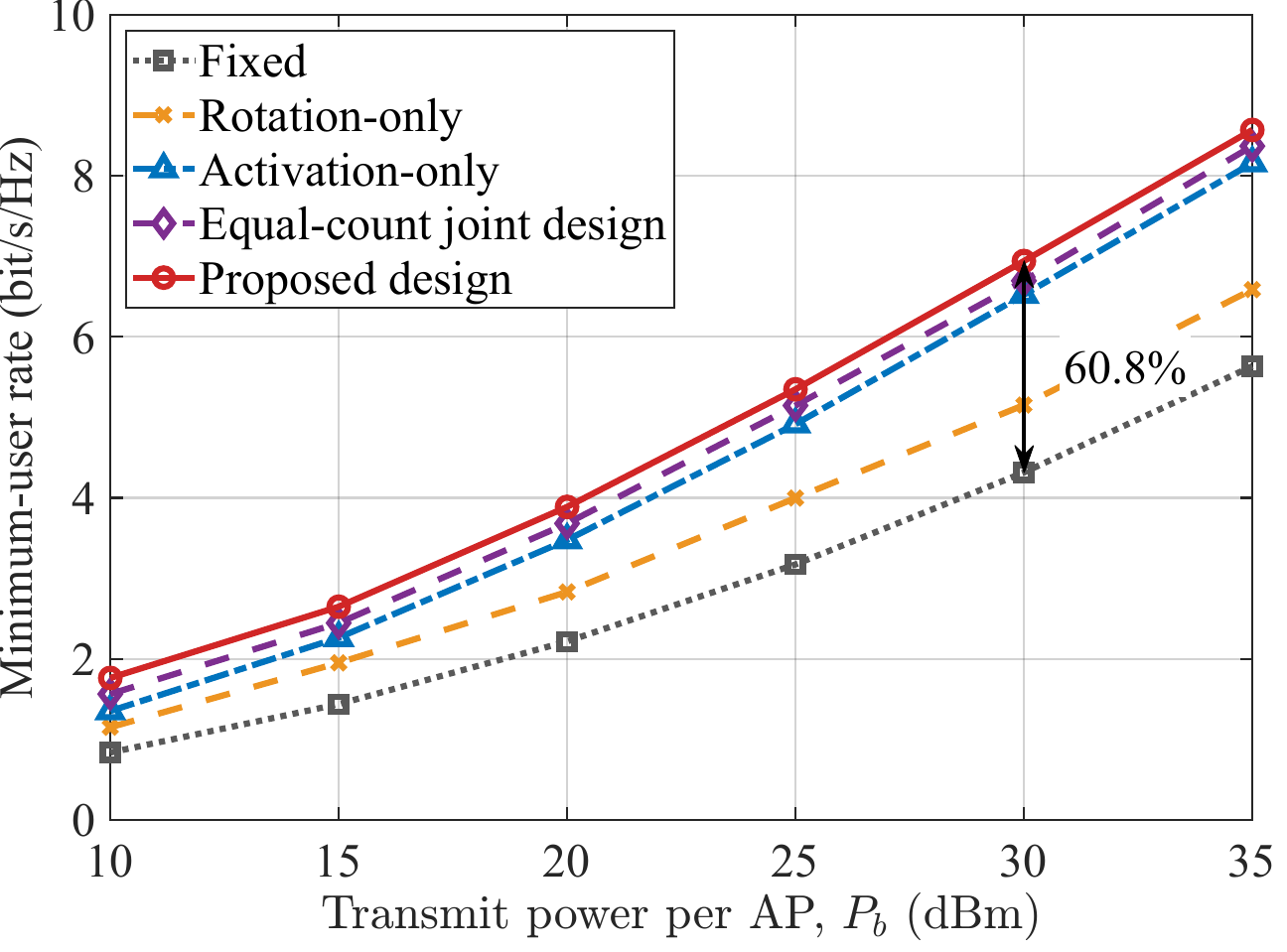}
	\caption{Minimum-user rate versus
		the per-AP transmit power.}
	\label{fig:rate_vs_power}
	\vspace{-4mm}
\end{figure}

Fig.~\ref{fig:rate_vs_power} shows the minimum-user
rate versus the per-AP transmit power $P_b$.
The proposed design consistently achieves the
highest rate, outperforming the fixed scheme
by $60.8\%$ at $P_b=30$~dBm.
This gain stems from jointly adjusting sparse
array apertures and panel orientations to
enhance channel strength and spatial separability.
Moreover, the activation-only scheme outperforms the rotation-only scheme,
as spatially dispersed users limit the ability
of a common panel orientation to simultaneously
enhance directional gains and suppress inter-user
channel correlation. In contrast, antenna
activation provides greater flexibility in
improving spatial separability by selecting
favorable antenna positions.
The further improvement over the equal-count
joint design highlights the benefit of adaptive
antenna allocation across APs.

\begin{figure}[t]
	\centering
	\includegraphics[width=0.4\textwidth]
	{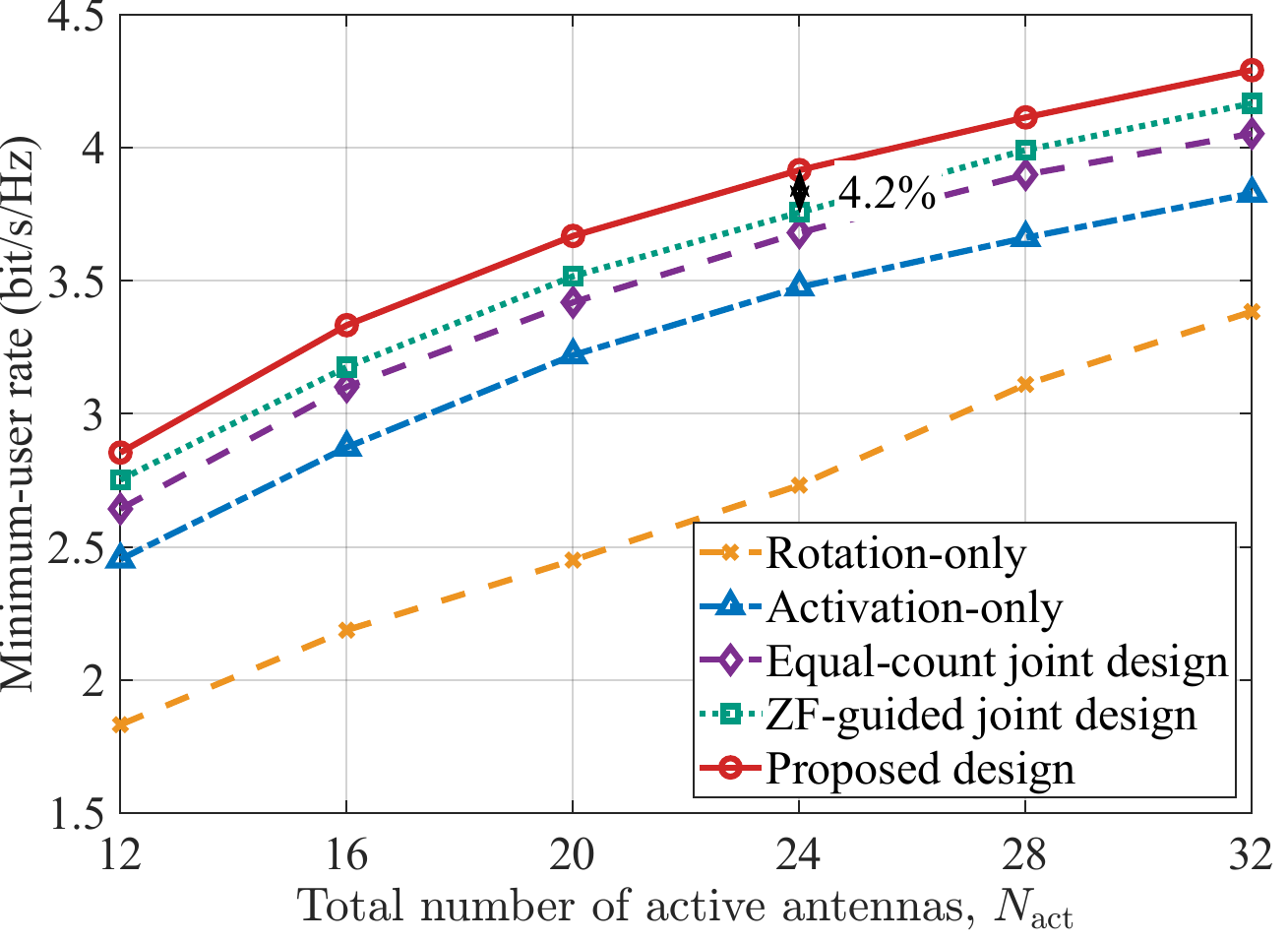}
	\caption{Minimum-user rate versus
		the total number of active antennas.}
	\label{fig:rate_vs_nact}
	\vspace{-7mm}
\end{figure}

Fig.~\ref{fig:rate_vs_nact} shows the minimum-user
rate versus the total number of active antennas.
To assess Proposition~\ref{prop:c_general_zf_allocation},
we introduce a ZF-guided joint design that fixes
AP-wise antenna counts based on the rounded
scenario-averaged allocation ratios from
\eqref{eq:c_optimal_zf_allocation_KB}, while jointly
optimizing activation positions and panel orientations.
The rate increases with $N_{\mathrm{act}}$
for all schemes, with diminishing marginal gains
due to the limited candidate aperture.
Notably, the proposed design achieves approximately
the same rate with $N_{\mathrm{act}}=20$
as activation-only scheme with $N_{\mathrm{act}}=28$.
The ZF-guided joint design consistently
outperforms the equal-count joint design,
validating the practical utility of the
analytical allocation for discrete activation.
At $N_{\mathrm{act}}=24$, the proposed design
further improves upon the ZF-guided joint
design by 4.2\%, demonstrating the benefit
of adaptive AP-wise antenna allocation.

\begin{figure}[t]
	\centering
		\includegraphics[width=0.4\textwidth]{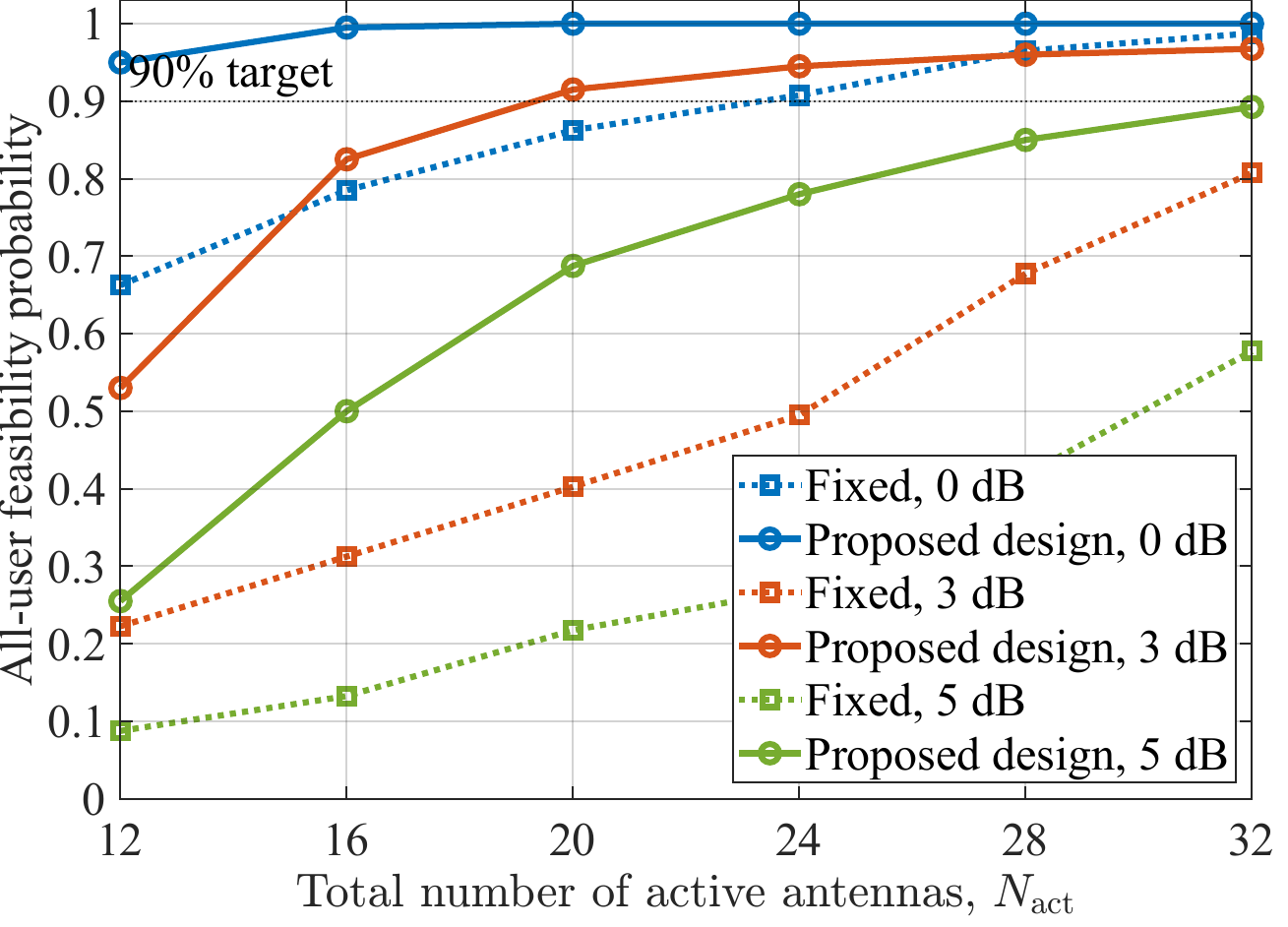}
	\caption{Empirical all-user feasibility probability versus the total
		number of active antennas under different SINR targets.}
	\label{fig:reliability_nact}
	\vspace{-4mm}
\end{figure}

Fig.~\ref{fig:reliability_nact} compares the empirical
all-user feasibility probabilities versus the total
number of active antennas under different SINR targets.
The feasibility probabilities are evaluated over $500$
independent nonempty user-channel scenarios.
A scenario is feasible if all users satisfy the
prescribed SINR target under optimized short-term
precoding. For both schemes, feasibility generally
increases with the antenna budget but decreases as
the SINR target becomes more stringent.
Notably, at a target of $3$~dB, the proposed design
achieves approximately $94.5\%$ feasibility with only
$N_{\mathrm{act}}=24$, exceeding the $90\%$ reliability
target, whereas the fixed scheme reaches only $80.5\%$
even with $N_{\mathrm{act}}=32$.
At the stringent $5$~dB target, the proposed scheme
achieves approximately $89.5\%$ feasibility at
$N_{\mathrm{act}}=32$, compared with $58\%$ for
the fixed scheme.

\begin{figure}[t]
	\centering
	\includegraphics[width=\columnwidth]
	{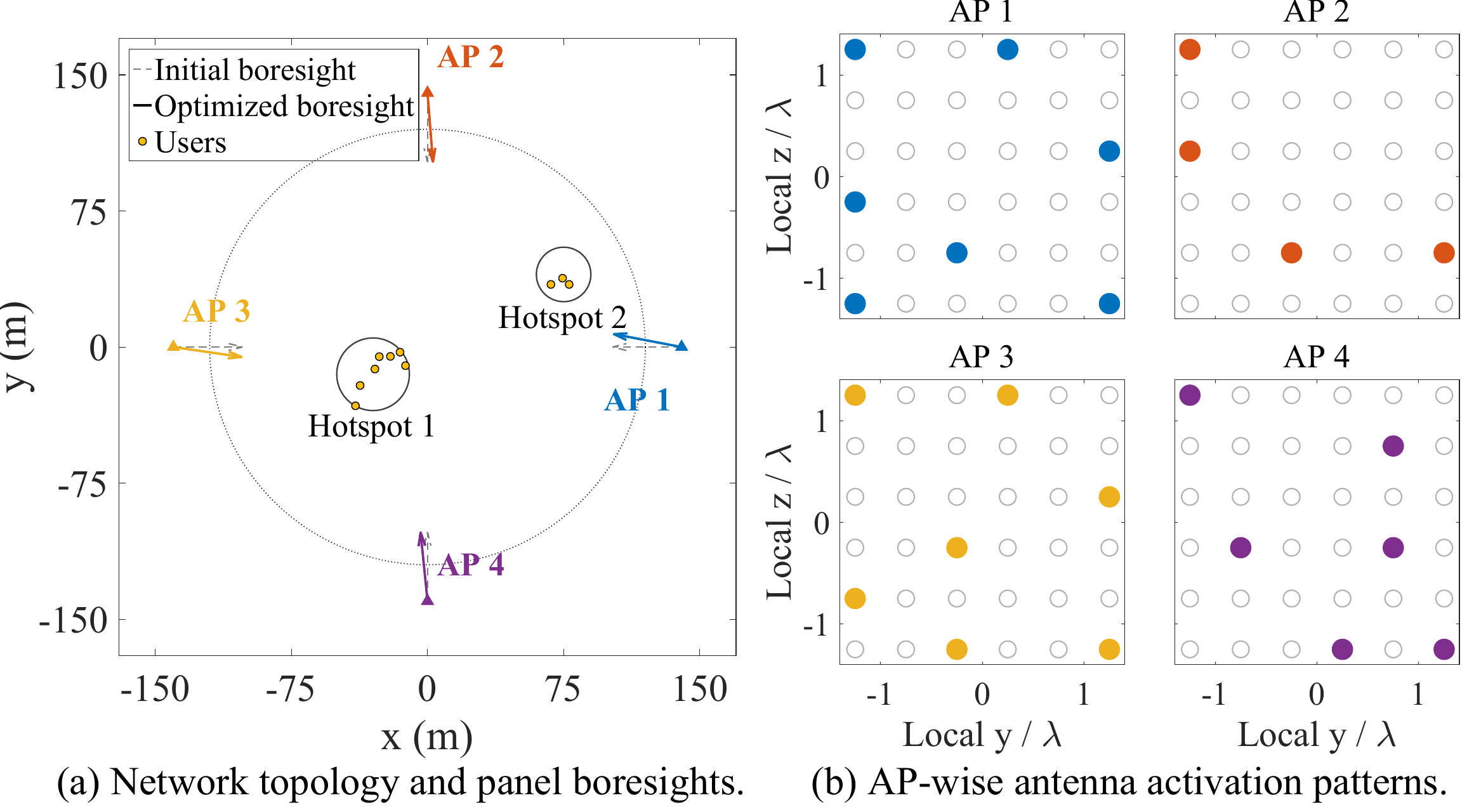}
	\includegraphics[width=\columnwidth]
	{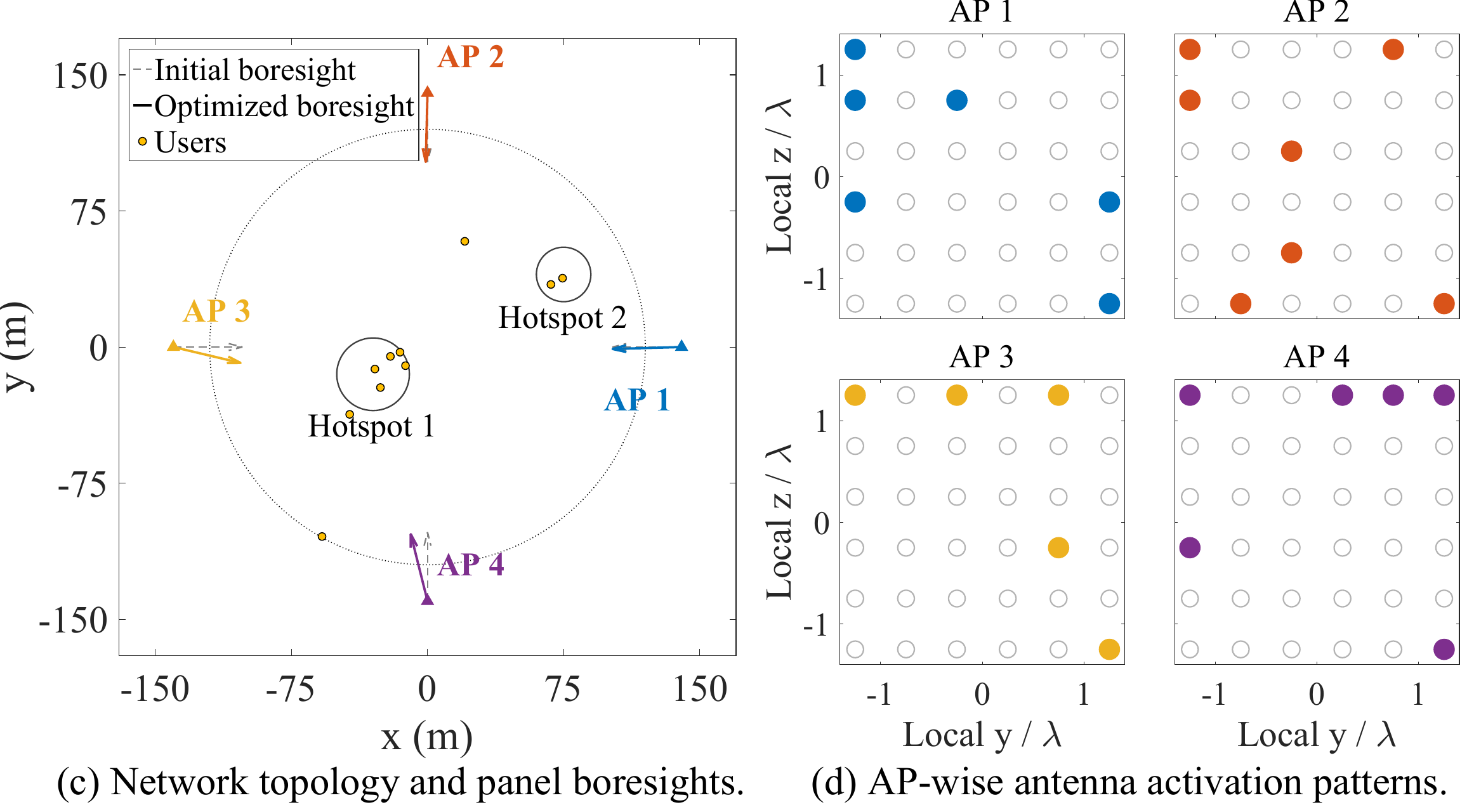}
\caption{Optimized long-term antenna configurations
	for $\xi=0$ (top) and $\xi=0.4$ (bottom).}
	\label{fig:optimized_configuration}
	\vspace{-7mm}
\end{figure}
Fig.~\ref{fig:optimized_configuration}
illustrates the optimized long-term antenna
configurations for $\xi=0$ and $\xi=0.4$.
As users become more dispersed, AP-wise
antenna allocation redistributes spatial
resources, local activation improves
multiuser separability, and panel rotation
adjusts directional gains and array geometry.
Together, these adaptations balance channel
strength and multiuser interference under
a fixed antenna budget.

\begin{figure}[t]
	\centering
	\includegraphics[width=0.4\textwidth]{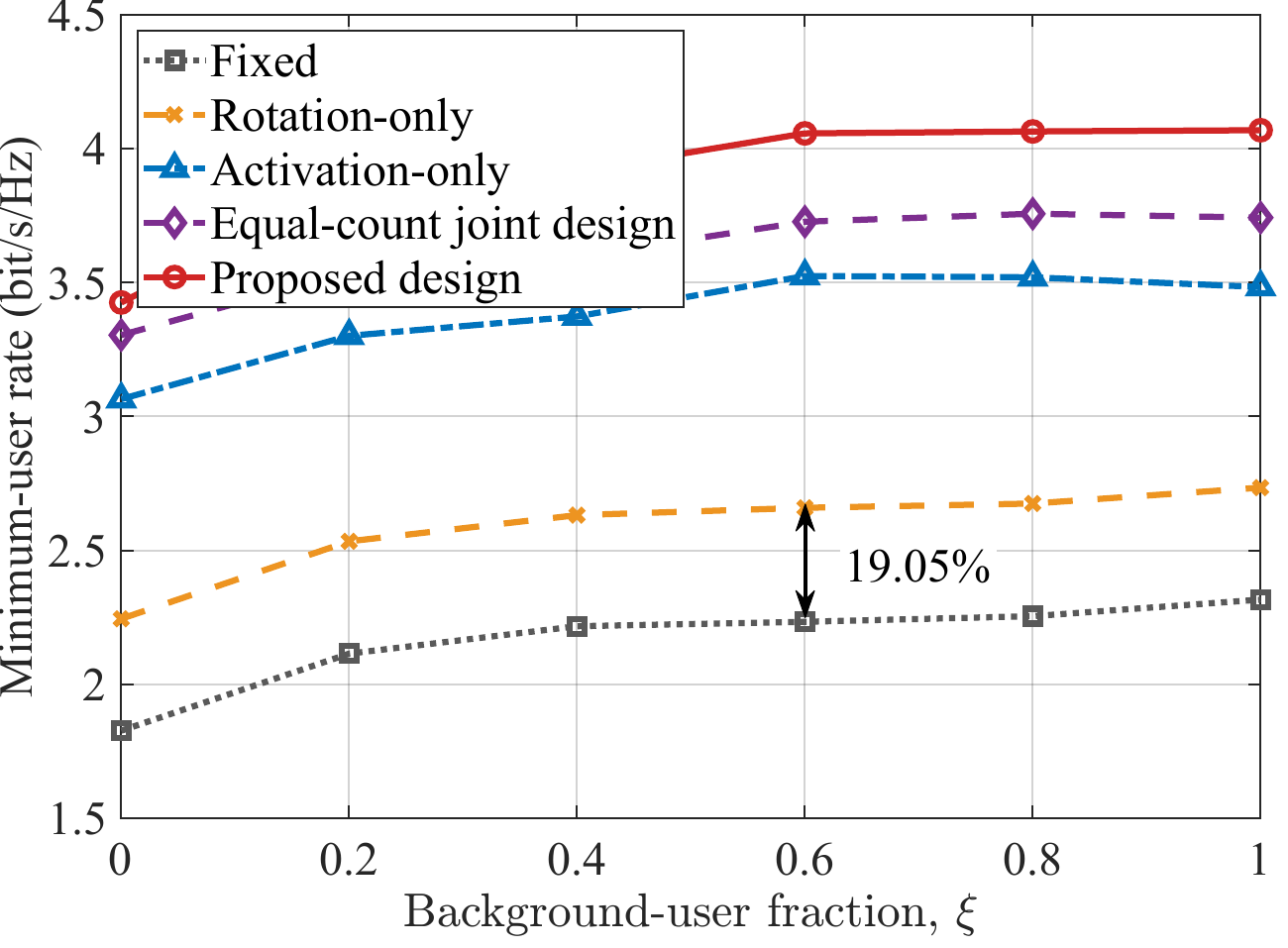}
\caption{Minimum-user rate versus the
	background-user fraction $\xi$.}
	\label{fig:rate_vs_xi}
	\vspace{-4mm}
\end{figure}

Fig.~\ref{fig:rate_vs_xi} investigates the impact of
user spatial dispersion by varying the background-user
fraction $\xi$ while keeping the mean user load fixed.
A smaller $\xi$ indicates stronger hotspot clustering,
whereas $\xi=1$ represents a spatially uniform
user distribution.
As $\xi$ increases, the minimum-user rates
generally improve due to enhanced spatial channel
separability and reduced inter-user interference,
with diminishing gains at larger $\xi$.
Compared with the fixed-array scheme, rotation-only
scheme achieves a $19.05\%$ rate improvement
at $\xi=0.6$, demonstrating the benefit of adapting
panel orientations to spatial user distributions.
 The activation-only scheme nevertheless outperforms the rotation-only design, while the proposed design achieves the highest rate for all considered $\xi$.

\begin{figure}[t]
	\centering
	\includegraphics[width=0.4\textwidth]
	{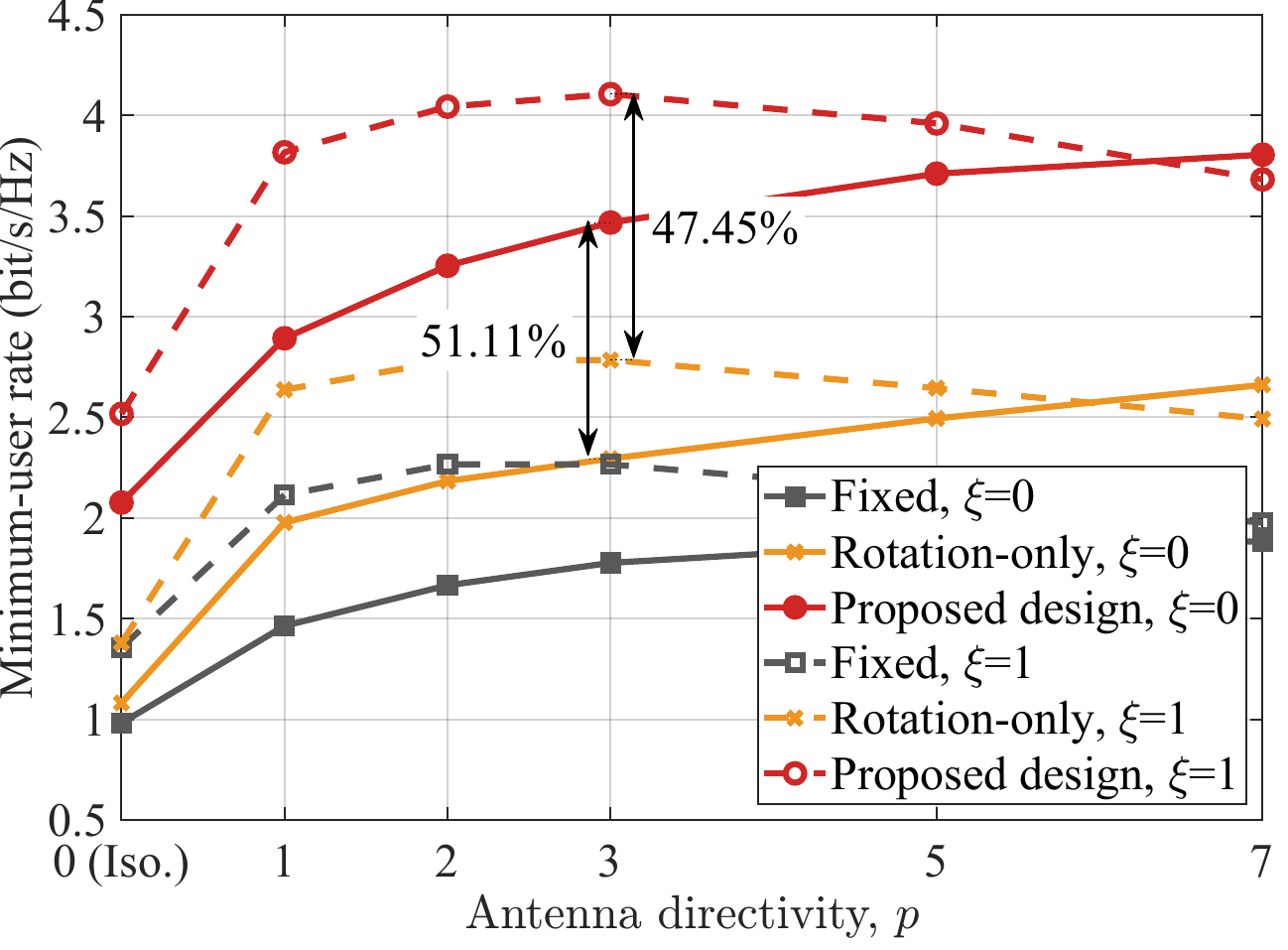}
	\caption{Minimum-user rate versus
		antenna directivity $p$.}
	\label{fig:rate_vs_p}
	\vspace{-7mm}
\end{figure}

Fig.~\ref{fig:rate_vs_p} studies the impact of
antenna directivity under different user spatial
distributions.
For $\xi=0$, the minimum-user rates generally
increase with $p$, as clustered users can better
exploit the enhanced directional gain.
In contrast, for $\xi=1$, the rates initially
improve but gradually decrease at larger $p$,
since narrower radiation patterns make it
more difficult to accommodate spatially
dispersed users.
These results reveal a tradeoff between directional
gain and spatial coverage, which depends on
the user distribution.
{At $p=3$, the proposed algorithm outperforms
rotation-only optimization by $51.11\%$ and
$47.45\%$ for $\xi=0$ and $\xi=1$, respectively,
showing additional gains from activation in both spatial distributions.}

\begin{figure}[t]
	\centering
	\includegraphics[width=0.4\textwidth]
	{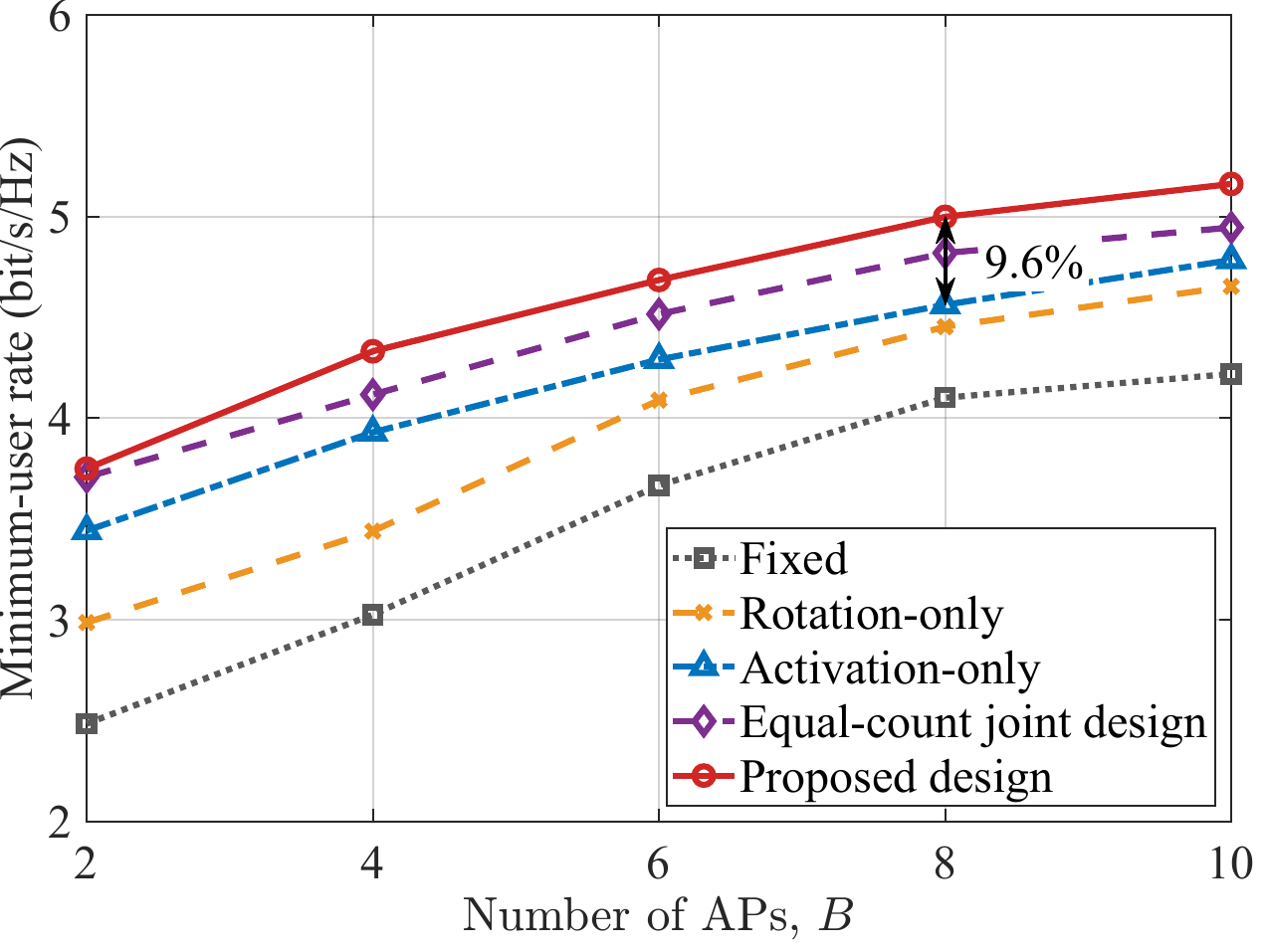}
\caption{{Minimum-user rate versus AP count.}}
\label{fig:rate_vs_B}
	\vspace{-7mm}
\end{figure}

Fig.~\ref{fig:rate_vs_B} investigates the impact
of the number of APs on the minimum-user
rate under a fixed network-wide activation budget
$N_{\mathrm{act}}=32$ and total transmit-power
budget $P_{\mathrm{tot}}=400$~mW.
Each AP has $36$ installed antennas and supports
up to $N_b^{\mathrm{RF}}=16$ active antennas,
with $P_b=P_{\mathrm{tot}}/B$.
When $N_{\mathrm{act}}$ is not divisible by $B$,
the equal-count baselines adopt a balanced integer
allocation, with AP-wise counts differing by at
most one.
As $B$ increases, all schemes achieve higher
minimum-user rates, with diminishing gains,
owing to enhanced macro-diversity and spatial
channel separability despite fixed network-wide active-antenna and transmit-power budgets.
At $B=8$, the proposed design achieves a $9.6\%$
higher rate than the activation-only scheme,
demonstrating the additional benefit of panel rotation.

\vspace{-2mm}
\section{Conclusion}
\label{Conclusion}
\vspace{-1mm}
This paper investigated an SRA-enabled cell-free downlink system with joint antenna activation and panel rotation. To exploit the spatial flexibility of SRAs while limiting hardware reconfiguration overhead, we formulated a two-timescale stochastic optimization problem to maximize the expected minimum-user rate. Under LoS propagation, our analysis revealed the complementary roles of antenna activation and panel rotation in enhancing channel strength and spatial separability. For two-user transmission, we established geometric conditions under which discrete antenna activation and panel rotation attain the continuous-position benchmark. For multi-AP transmission, the derived optimal ZF SINR and closed-form antenna allocation demonstrated the importance of distributed channel separation in allocating limited antenna resources. To solve the mixed-integer nonconvex problem, we developed an SAA-based alternating algorithm integrating UL-DL duality, WMMSE-based orientation optimization, and multi-start antenna activation search. Numerical results demonstrated that the proposed design enhances multiuser fairness and antenna resource utilization through joint aperture and orientation reconfiguration.

\appendices
	\refstepcounter{section}
\vspace{-2mm}
\section*{APPENDIX~\Alph{section}: Proof of Lemma~\ref{lem:twouser_sinr}}
\label{app:b_two_user_sinr}
\vspace{-2mm}
Let $\mathbf u_k=\mathbf g_k^*/\sigma_k$ and
$a_k=\|\mathbf u_k\|_2^2=s_k/\sigma_k^2$.
For a target SINR $t>0$, we consider the minimum
total transmit power required to satisfy both
{users' SINR constraints. The dual is}
\begin{align}
	\max_{\lambda_1,\lambda_2\geq 0}
	\quad &\lambda_1+\lambda_2 \notag\\
	\mathrm{s.t.}\quad
	&\mathbf I+\lambda_j\mathbf u_j\mathbf u_j^{\mathrm H}
	-{\lambda_k}/{t}
	\mathbf u_k\mathbf u_k^{\mathrm H}
	\succeq\mathbf0, j\neq k.
\end{align}
At the optimal dual point, the semidefinite
constraints are active in the corresponding
rank-one directions. Applying the
Sherman--Morrison identity and defining
$z_k=\lambda_k a_k$ yield
\begin{align}
	z_1=\frac{t(1+z_2)}
	{1+(1-r)z_2},
	z_2=\frac{t(1+z_1)}
	{1+(1-r)z_1}.
\end{align}
Subtracting the cross-multiplied equations gives
$(1+t)(z_1-z_2)=0$, and hence $z_1=z_2=z$.
Substitution yields
$(1-r)z^2+(1-t)z-t=0.
$
By strong duality, the minimum required transmit
power is
\begin{align}
	P_{\min}(t)
	&= \lambda_1+\lambda_2
	= z\left({1}/{a_1}+{1}/{a_2}\right).
\end{align}
At the optimal max-min SINR, the power budget is
attained, yielding
$z=P/(1/a_1+1/a_2)=x$.
Substituting this into the quadratic equation and
solving for $t$ yields
$\tau_2^\star=x-rx^2/(1+x)$,
which proves \eqref{eq:b_two_user_sinr}.
Finally, differentiation gives
\begin{align}
	\frac{\partial T}{\partial x}
	=1-r+\frac{r}{(1+x)^2}>0,
	\frac{\partial T}{\partial r}
	=-\frac{x^2}{1+x}<0,
\end{align}
for $x>0$ and $0\leq r\leq1$.

\refstepcounter{section}
\vspace{-3mm}
\section*{APPENDIX~\Alph{section}: Proof of Proposition~\ref{prop:c_general_zf_allocation}}
\label{app:c_zf_allocation}
\vspace{-1mm}

Let $\mathbf W_b=[\mathbf w_{b,1},\ldots,
\mathbf w_{b,K}]$ denote the precoding matrix
at AP $b$. We factor the reference Gram matrix
as $\mathbf A_b=\overline{\mathbf H}_b
\overline{\mathbf H}_b^{\mathrm H}$ and express
the equivalent local channel as
$\mathbf H_b(n_b)=\sqrt{n_b}\,\overline{\mathbf H}_b$.
The virtual channel
$
\widetilde{\mathbf H}
=[\sqrt{NP_1}\,\overline{\mathbf H}_1,\ldots,
\sqrt{NP_B}\,\overline{\mathbf H}_B]$
has Gram matrix
$\widetilde{\mathbf H}\widetilde{\mathbf H}^{\mathrm H}
=N\mathbf M$. By \eqref{eq:c_multiuser_sum_power_value},
its optimal max-min SINR under unit sum power is
$\mathcal T_K(N\mathbf M,\mathbf D)$.
Let $\widetilde{\mathbf V}_b$ denote the AP-$b$ block
of the virtual precoding matrix $\widetilde{\mathbf V}$.
For any feasible counts and
AP precoders, we set
$\widetilde{\mathbf V}_b=\sqrt{n_b/(NP_b)}\,\mathbf W_b$. Then, we have
\begin{equation}
	\sum\nolimits_b\|\widetilde{\mathbf V}_b\|_{\mathrm F}^2\le1,
	\widetilde{\mathbf H}\widetilde{\mathbf V}
	=\sum\nolimits_b\mathbf H_b(n_b)\mathbf W_b.
	\label{eq:c_appendix_virtual_mapping}
\end{equation}
Thus, the virtual sum-power problem provides an upper bound.
Conversely, for a unit-power optimal virtual precoder,
let $q_b=\|\widetilde{\mathbf V}_b^\star\|_{\mathrm F}^2$ and choose
\[
n_b=Nq_b,
\mathbf W_b=\sqrt{P_b/q_b}\,\widetilde{\mathbf V}_b^\star
\quad(q_b>0).
\]
For $q_b=0$, set $n_b=0$ and $\mathbf W_b=\mathbf0$.
These choices satisfy every AP power constraint and preserve
the received signals. The upper bound is therefore attained,
which proves \eqref{eq:c_optimal_multiuser_ap_sinr} and gives
the optimal count construction. Both transformations also
preserve ZF, so the same equivalence holds when precoding
is restricted to ZF.

Suppose $\mathbf M\succ\mathbf0$
and let $s=\operatorname{tr}(\mathbf D\mathbf M^{-1})$.
For target SINR $\tau$, the minimum-norm virtual ZF precoder is
\begin{equation}
	\widetilde{\mathbf V}_{\mathrm{ZF}}(\tau)
	=\widetilde{\mathbf H}^{\mathrm H}(N\mathbf M)^{-1}
	\sqrt{\tau}\,\mathbf D^{1/2},
	\label{eq:c_appendix_virtual_zf}
\end{equation}
where the desired signal phases are chosen to be zero.
Its power is $\tau s/N$, so unit sum power gives
$\tau_{\mathrm{ZF,c}}^\star=N/s$. The power of its $b$th AP
block is
\[
\|\widetilde{\mathbf V}_{b,\mathrm{ZF}}(\tau)\|_{\mathrm F}^2
=\frac{\tau P_b}{N}
\operatorname{tr}(\mathbf D\mathbf M^{-1}
\mathbf A_b\mathbf M^{-1})
=\frac{\tau P_bT_b}{N}.
\]
At $\tau=N/s$, the preceding count construction therefore
gives $n_{b,\mathrm c}^\star=NP_bT_b/s$, proving
\eqref{eq:c_optimal_zf_allocation_KB} and
\eqref{eq:c_general_optimal_zf_sinr}.

The optimal allocation is unique. To prove this, we consider
any feasible ZF design attaining $\tau=N/s$ and its
corresponding virtual precoder $\widetilde{\mathbf V}$. Since the
minimum virtual transmit power required for this SINR
is $\tau s/N=1$, $\widetilde{\mathbf V}$ must have unit power and
attain this minimum. It therefore coincides with
\eqref{eq:c_appendix_virtual_zf}, up to the phases of
the individual streams. These phases do not affect
the AP block powers, so
$
\|\widetilde{\mathbf V}_b\|_{\mathrm F}^2=\frac{P_bT_b}{s}$.
On the other hand, the per-AP power constraints imply
$\|\widetilde{\mathbf V}_b\|_{\mathrm F}^2\le n_b/N$.
Because $\sum_b\|\widetilde{\mathbf V}_b\|_{\mathrm F}^2
=\sum_b n_b/N=1$, equality must hold for every $b$.
Hence $n_b=NP_bT_b/s$, establishing uniqueness.
If $\mathbf M$ is singular, there exists a nonzero
vector $\mathbf v$ such that $\mathbf A_b\mathbf v=\mathbf0$
for every $b$, since $P_b>0$ and $\mathbf A_b\succeq\mathbf0$.
Thus, $\sum_b n_b\mathbf A_b$ is singular for every
allocation, and ZF cannot provide a positive SINR
to all users.

\bibliographystyle{IEEEtran}
\vspace{-3mm}
\begin{spacing}{0.95}
	\bibliography{thesis}

@IEEEtranBSTCTL{reference:BSTcontrol,
	CTLuse_article_number = "yes",
	CTLuse_paper = "yes",
	CTLuse_forced_etal = "no",
	CTLmax_names_forced_etal = "10",
	CTLnames_show_etal = "1"
}

@book{balanis2016antenna,
	author    = {Constantine A. Balanis},
	title     = {Antenna Theory: Analysis and Design},
	edition   = {4},
	publisher = {John Wiley \& Sons},
	year      = {2016}
}

@ARTICLE{Papazafeiropoulos2021EE,
	author={Papazafeiropoulos, Anastasios and others},
	journal={IEEE Trans. Green Commun. Netw.}, 
	title={Towards Optimal Energy Efficiency in Cell-Free Massive {MIMO} Systems}, 
	year={Feb. 2021},
	volume={5},
	number={2},
	pages={816-831},
	doi={10.1109/TGCN.2021.3059206}}

@ARTICLE{Feng2022SE,
	author={Feng, Chenghao and others},
	journal={IEEE Trans. Wireless Commun.}, 
	title={Weighted Sum Rate Maximization of the mmWave Cell-Free {MIMO} Downlink Relying on Hybrid Precoding}, 
	year={Sept. 2021},
	volume={21},
	number={4},
	pages={2547-2560},
	doi={10.1109/TWC.2021.3113458}}

@ARTICLE{Wang6G2023,
	author={Wang, Cheng-Xiang and others},
	journal={IEEE Commun. Surveys Tuts.}, 
	title={On the Road to {6G}: Visions, Requirements, Key Technologies, and Testbeds}, 
	year={Feb. 2023},
	volume={25},
	number={2},
	pages={905-974},
	doi={10.1109/COMST.2023.3249835}}

@article{ngo2017cellfree,
	author = {Ngo, Hien Quoc and Ashikhmin, Alexei and Yang, Hong and Larsson, Erik G. and Marzetta, Thomas L.},
	title = {Cell-Free Massive {MIMO} Versus Small Cells},
	journal = {IEEE Trans. Wireless Commun.},
	year = {Jan. 2017}, volume = {16}, number = {3}, pages = {1834--1850},
	doi = {10.1109/TWC.2017.2655515}
}

@article{demir2021foundations,
	author  = {Demir, {\"O}zlem Tu{\u{g}}fe and others},
	title   = {Foundations of User-Centric Cell-Free Massive {MIMO}},
	journal = {Found. Trends Signal Process.},
	volume  = {14},
	number  = {3--4},
	pages   = {162--472},
	year    = {2021},
	doi     = {10.1561/2000000109}
}

@article{miretti2022team,
	author  = {Miretti, Lorenzo and Bj{\"o}rnson, Emil and Gesbert, David},
	title   = {Team {MMSE} Precoding With Applications to Cell-Free Massive {MIMO}},
	journal = {IEEE Trans. Wireless Commun.},
	volume  = {21},
	number  = {8},
	pages   = {6242--6255},
	year    = {Feb. 2022}
}

@article{interdonato2019ubiquitous,
	author = {Interdonato, Giovanni and Bj{\"o}rnson, Emil and Ngo, Hien Quoc and Frenger, P{\aa}l and Larsson, Erik G.},
	title = {Ubiquitous Cell-Free Massive {MIMO} Communications},
	journal = {EURASIP J. Wirel. Commun. Netw.},
	year = {Aug. 2019}, volume = {2019}, pages = {197},
	doi = {10.1186/s13638-019-1507-0}
}

@article{nguyen2022rfactivation,
	author = {Nguyen, Nhan Thanh and others},
	title = {Hybrid Beamforming and Adaptive {RF} Chain Activation for Uplink Cell-Free Millimeter-Wave Massive {MIMO} Systems},
	journal = {IEEE Trans. Veh. Technol.},
	year = {May 2022}, volume = {71}, number = {8}, pages = {8739--8755},
	doi = {10.1109/TVT.2022.3176389}
}

@article{wang2026antennaselection,
	author = {Wang, Liangzhi and Chen, Chen and Zhang, Jie and Fischione, Carlo},
	title = {Learning-Based Joint Antenna Selection and Precoding Design for Cell-Free {MIMO} Networks},
	journal = {IEEE Trans. Veh. Technol.},
	year = {Jul. 2025}, volume = {75}, number = {1}, pages = {579--592},
	doi = {10.1109/TVT.2025.3590425}
}

@misc{zheng2026sra,
	author = {Zheng, Ailing and Wu, Qingqing and Liu, Xiyuan and Chen, Wen},
	title = {Sparse Rotatable Arrays ({SRA}): Unifying Array Aperture and Antenna Directivity for Wireless Communications},
	year = {early access, 2026}, eprint = {2608.11666}, archivePrefix = {arXiv},
	url = {https://arxiv.org/abs/2608.11666}
}

@article{shi2025cellfree6dma,
	author = {Shi, Xiaoming and Shao, Xiaodan and Zheng, Beixiong and Zhang, Rui},
	title = {{6DMA}-Aided Cell-Free Massive {MIMO} Communication},
	journal = {IEEE Wireless Commun. Lett.},
	year = {Feb. 2025}, volume = {14}, number = {5}, pages = {1361--1365},
	doi = {10.1109/LWC.2025.3542546}
}

@article{pi2026multiap6dma,
	author = {Pi, Xiangyu and others},
	title = {{6D} Movable Antenna Enhanced Multi-Access Point Coordination via Position and Orientation Optimization},
	journal = {IEEE Trans. Wireless Commun.},
	year = {Jul. 2025}, volume = {25}, pages = {915--930},
	doi = {10.1109/TWC.2025.3587803}
}

@article{peng2026racellfree,
	author = {Peng, Xingxiang and Wu, Qingqing and Zheng, Ziyuan and Zhu, Yanze and Chen, Wen and Huang, Penghui and Gao, Ying and Wang, Honghao},
	title = {Cell-Free {MIMO} With Rotatable Antennas: When Macro-Diversity Meets Antenna Directivity},
	journal = {IEEE Trans. Wireless Commun.},
	year = {Jul. 2026}, volume = {25}, pages = {20867--20882},
	doi = {10.1109/TWC.2026.3716707}
}

@ARTICLE{pan2025racellfree,
	author={Pan, Kecheng and others},
	journal={IEEE Wireless Commun. Lett.}, 
	title={Rotatable Antenna-Enhanced Cell-Free Communication}, 
	year={May 2026},
	volume={15},
	number={},
	pages={3458-3462},
	doi={10.1109/LWC.2026.3696844}}

@misc{zhang2026twotimescale6dma,
	author = {Zhang, Yichi and others},
	title = {{6D} Movable Antenna Enhanced Cell-free {MIMO}: Two-timescale Decentralized Beamforming and Antenna Movement Optimization},
	year = {early access, 2026}, eprint = {2601.04969}, archivePrefix = {arXiv},
	url = {https://arxiv.org/abs/2601.04969}
}

@article{farooq2021utility,
	author  = {M. Farooq and H. Q. Ngo and E. K. Hong
	and L. N. Tran},
	title   = {Utility Maximization for Large-Scale
	Cell-Free Massive {MIMO} Downlink},
	journal = {IEEE Trans. Commun.},
	volume  = {69},
	number  = {10},
	pages   = {7050--7062},
	year    = {2021},
	doi     = {10.1109/TCOMM.2021.3097718}
}

@ARTICLE{YuLan2007,
	author={Yu, Wei and Lan, Tian},
	journal={IEEE Trans. Signal Process.}, 
	title={Transmitter Optimization for the Multi-Antenna Downlink With Per-Antenna Power Constraints}, 
	year={Jun. 2007},
	volume={55},
	number={6},
	pages={2646-2660},
	doi={10.1109/TSP.2006.890905}}

@ARTICLE{CaiEtAl2012,
	author={Cai, Desmond W. H. and Quek, Tony Q. S. and Tan, Chee Wei and Low, Steven H.},
	journal={IEEE Trans. Signal Process.}, 
	title={Max-Min {SINR} Coordinated Multipoint Downlink Transmission-Duality and Algorithms}, 
	year={Jul. 2012},
	volume={60},
	number={10},
	pages={5384-5395},
	doi={10.1109/TSP.2012.2208631}}

@ARTICLE{Shao20256DMA,
	author={Shao, Xiaodan and Jiang, Qijun and Zhang, Rui},
	journal={IEEE Trans. Wireless Commun.}, 
	title={{6D} Movable Antenna Based on User Distribution: Modeling and Optimization}, 
	year={Nov. 2024},
	volume={24},
	number={1},
	pages={355-370},
	doi={10.1109/TWC.2024.3492195}}

@ARTICLE{Wong2020,
	author={Wong, Kai-Kit and others},
	journal={IEEE Trans. Wireless Commun.}, 
	title={Fluid Antenna Systems}, 
	year={Nov. 2020},
	volume={20},
	number={3},
	pages={1950-1962},
	doi={10.1109/TWC.2020.3037595}}

@ARTICLE{ZhengRA2026,
	author={Zheng, Beixiong and Wu, Qingjie and Ma, Tiantian and Zhang, Rui},
	journal={IEEE Trans. Commun.}, 
	title={Rotatable Antenna-Enabled Wireless Communication: Modeling and Optimization}, 
	year={Mar. 2026},
	volume={74},
	number={},
	pages={6825-6842},
	doi={10.1109/TCOMM.2026.3672230}}

@ARTICLE{Zhu2024MA,
	author={Zhu, Lipeng and Ma, Wenyan and Zhang, Rui},
	journal={IEEE Trans. Wireless Commun.}, 
	title={Modeling and Performance Analysis for Movable Antenna Enabled Wireless Communications}, 
	year={Nov. 2023},
	volume={23},
	number={6},
	pages={6234-6250},
	doi={10.1109/TWC.2023.3330887}}

@ARTICLE{AilngRASecure2026,
	author={Zheng, Ailing and Wu, Qingqing and Peng, Xingxiang and Peng, Qiaoyan and Zheng, Ziyuan and Chen, Yuxuan and Chen, Wen},
	journal={IEEE Trans. Wireless Commun.}, 
	title={Rotatable Antenna-Aided Secure {MIMO} Communications: Joint Transmit and Receive Orientation Design}, 
	year={Sept. 2026},
	volume={25},
	number={},
	pages={23864-23881},
	doi={10.1109/TWC.2026.3735258}}

@ARTICLE{Zhu2024CM,
	author={Zhu, Lipeng and Ma, Wenyan and Zhang, Rui},
	journal={IEEE Commun. Mag.}, 
	title={Movable Antennas for Wireless Communication: Opportunities and Challenges}, 
	year={Oct. 2023},
	volume={62},
	number={6},
	pages={114-120},
	doi={10.1109/MCOM.001.2300212}}

@ARTICLE{Shao2025-1,
	author={Shao, Xiaodan and Zhang, Rui},
	journal={IEEE Commun. Mag.}, 
	title={{6DMA} Enhanced Wireless Network with Flexible Antenna Position and Rotation: Opportunities and Challenges}, 
	year={Mar. 2025},
	volume={63},
	number={4},
	pages={121-128},
	doi={10.1109/MCOM.002.2400333}}

@ARTICLE{Zheng2026,
		author={Zheng, Ailing and Wu, Qingqing and Zheng, Ziyuan and Peng, Qiaoyan and Zhu, Yanze and Wang, Honghao and Chen, Wen and Zhang, Guoying},
		journal={IEEE Trans. Commun.}, 
		title={Wireless Communication With Cross-Linked Rotatable Antenna Array: Architecture Design and Rotation Optimization}, 
		year={Jun. 2026},
		volume={74},
		number={},
		pages={10261-10276},
		doi={10.1109/TCOMM.2026.3704189}}
\end{spacing}

\end{document}